\documentclass{pasa}%

\title[Lessons from MWA Polarimetry]{The challenges of low-frequency radio polarimetry: lessons from the Murchison Widefield Array}

\author[Lenc et al.]{E.~Lenc$^{1,2}$,
C.~S.~Anderson$^{3}$,
N.~Barry$^{4}$,
J.~D.~Bowman$^{5}$,
I.~H.~Cairns$^{6}$,
J.~S.~Farnes$^{7}$,
B.~M.~Gaensler$^{1,2,8}$,
G.~Heald$^{3}$,
M.~Johnston-Hollitt$^{9,10}$,
D.~L.~Kaplan$^{11}$,
C.~R.~Lynch$^{1,2}$,
P.~I.~McCauley$^{6}$,
D.~A.~Mitchell$^{12,2}$,
J.~Morgan$^{13,2}$,
M.F.~Morales$^{4}$,
Tara~Murphy$^{1,2}$,
A.~R.~Offringa$^{14}$,
S.~M.~Ord$^{12,2}$,
B.~Pindor$^{15,2}$,
C.~Riseley$^{3}$,
E.~M.~Sadler$^{1,2}$,
C.~Sobey$^{3,13}$,
M.~Sokolowski$^{13,2}$,
I.~S.~Sullivan$^{4}$,
S.~P.~O'Sullivan$^{16}$,
X.~H.~Sun$^{17}$,
S.~E.~Tremblay$^{13,2}$,
C.~M.~Trott$^{13,2}$,
R.~B.~Wayth$^{13,2}$\\
\affil{$^{1}$~Sydney Institute for Astronomy, School of Physics, The University of Sydney, NSW 2006, Australia}
\affil{$^{2}$~ARC Centre of Excellence for All-sky  Astrophysics (CAASTRO)}
\affil{$^{3}$~CSIRO Astronomy and Space Science (CASS), 26 Dick Perry Ave, Kensington, WA 6151, Australia}
\affil{$^{4}$~Department of Physics, University of Washington, Seattle, WA 98195, USA}
\affil{$^{5}$~School of Earth and Space Exploration, Arizona  State University, Tempe, AZ 85287, USA}
\affil{$^{6}$~School of Physics, The University of Sydney, NSW 2006, Australia}
\affil{$^{7}$~Department of Astrophysics/IMAPP, Radboud University, PO Box 9010, NL-6500 GL Nijmegen, The Netherlands}
\affil{$^{8}$~Dunlap Institute for Astronomy and Astrophysics, University of Toronto, 50 St.\ George Street, Toronto, ON M5S 3H4, Canada}
\affil{$^{9}$~School of Chemical \& Physical Sciences,  Victoria University of Wellington, Wellington 6140, New Zealand}
\affil{$^{10}$~Peripety Scientific Ltd., PO Box 11355 Manners Street, Wellington 6142, New Zealand}
\affil{$^{11}$~Department of Physics, University of Wisconsin--Milwaukee, Milwaukee, WI 53201, USA}
\affil{$^{12}$~CSIRO Astronomy and Space Science (CASS), PO Box  76, Epping, NSW 1710, Australia}
\affil{$^{13}$~International Centre for Radio Astronomy Research, Curtin University, Bentley, WA 6102, Australia}
\affil{$^{14}$~ASTRON, The Netherlands Institute for Radio Astronomy, Postbus 2, 7990 AA, Dwingeloo, The Netherlands}
\affil{$^{15}$~School of Physics, The University of Melbourne, Parkville, VIC 3010, Australia}
\affil{$^{16}$~Instituto de Astronom\'{i}a, Universidad Nacional Aut\'{o}noma de M\'{e}xico (UNAM), A.P. 70-264, 04510 M\'{e}xico, D.F., Mexico.}
\affil{$^{17}$~School of Physics and Astronomy, Yunnan University, Kunming, 650500, China}
}

\jid{PASA}
\doi{10.1017/pas.\the\year.xxx}
\jyear{\the\year}

\usepackage[authoryear]{natbib}
\bibpunct{(}{)}{;}{a}{}{,}
\usepackage{aas_macros}
\usepackage{hyperref} 
\usepackage{bm} 

\usepackage{microtype}
\hypersetup{colorlinks,citecolor=blue,linkcolor=blue,urlcolor=blue}

\newcommand\arcdeg{\mbox{$^\circ$}}%
\newcommand\arcmin{\mbox{$^\prime$}}%
\newcommand\arcsec{\mbox{$^{\prime\prime}$}}%

\begin{document}%

\begin{abstract}
We present techniques developed to calibrate and correct Murchison Widefield Array (MWA) low frequency ($72-300$\,MHz) radio observations for polarimetry.  The extremely wide field-of-view, excellent instantaneous $(u,v)$-coverage and sensitivity to degree-scale structure that the MWA provides enable instrumental calibration, removal of instrumental artefacts, and correction for ionospheric Faraday rotation through imaging techniques. With the demonstrated polarimetric capabilities of the MWA, we discuss future directions for polarimetric science at low frequencies to answer outstanding questions relating to polarised source counts, source depolarisation, pulsar science, low-mass stars, exoplanets, the nature of the interstellar and intergalactic media, and the solar environment.

\end{abstract}
\begin{keywords}
polarisation -- techniques: polarimetric -- radio continuum: general 
\end{keywords}
\maketitle%
\section{Introduction}
\label{sec:introduction}

Polarised radio emission arises generically from processes involving magnetised plasmas which are ubiquitous in the Universe.  For example, polarisation appears, at some level, in all sources of synchrotron radiation.  Measuring and studying the polarised emission (polarimetry) can provide insight into physical processes occurring in systems that range from our own atmosphere (the ionosphere), the intervening interstellar medium, and out to high-redshift galaxies. Science that depends on polarisation is part of the scientific motivation for a number of low frequency instruments, however, doing accurate polarimetry at low frequencies is challenging.

Polarised sources generally depolarise with increasing wavelength as a result of Faraday depolarisation \citep{Farnsworth:2011,deBruyn:2012,giessubel:2013,Anderson:2015} during propagation from the radio source. Relevant regions include the extragalactic, galactic, interstellar, and interplanetary plasmas. When linearly polarised radiation propagates through a magnetised plasma it undergoes Faraday rotation according to:
\begin{equation}
\chi(\lambda^{2}) = \chi_{0} + \phi\lambda^{2} = \chi_{0} + \int_{0}^{z} dz'\ RM(z')\ \lambda^{2}
\end{equation}
where $\chi(\lambda^{2})$ is the measured linear polarisation angle (rad) at wavelength $\lambda$ (m), $\chi_{0}$ is the intrinsic polarisation angle (rad), $\phi$ is the Faraday depth (rad\,m$^{-2}$), and the final term is the integral along the path from the source ($z' = 0$) to the observer ($z' = z$) of the rotation measure $RM(z')$, which depends on the spatially varying plasma density and component of the magnetic field along the path. Faraday rotation is particularly pronounced at long wavelengths because of the wavelength-squared dependence. The ionosphere introduces additional Faraday rotation which, if not corrected for, can further depolarise the signal. The fact that polarized signals are generally only a few percent as strong as the total intensity of the source itself means that measuring polarisation is challenging.

Polarimetry programs are underway on a range of low frequency instruments including the LOw-Frequency ARray \citep[LOFAR,][]{vanHaarlem:2013}, the Long Wavelength Array \citep[LWA,][]{Ellingson:2009}, the Precision Array for Probing the Epoch of Reionization \citep[PAPER,][]{Parsons:2010}, and the Murchison Widefield Array \citep[MWA,][]{Tingay:2013}. 
In addition, the ongoing upgrade to the Giant Metrewave Radio Telescope \citep[GMRT,][]{Swarup:1990} will allow it to make sensitive polarimetric measurements. Processing 
polarimetric data pushes instruments to their software and hardware limits, since it is not only processor and data intensive, but also highly sensitive to instrumental and calibration errors.
Hence polarimetry is a powerful diagnostic tool when evaluating and characterizing the performance of a new instrument \citep{Sutinjo:2015v50p52S}.

In this paper, we address some of the key technical challenges of doing polarimetry at low frequencies with the MWA. Section \ref{sec:mwapol} outlines some of the early science results obtained with the MWA; Section \ref{sec:mwasw} looks at software tools used to process MWA data; Section \ref{sec:mwacal} summarizes calibration methods that were developed for MWA processing; and Section \ref{sec:science} considers science that could be explored given our understanding of the MWA capabilities as it stands today and in the near future.

\section{Polarimetry with the MWA}
\label{sec:mwapol}

The MWA is located at the Murchison Radio Observatory in Western Australia. It consists of an array of up to 128 connected tiles where each tile is comprised of a regular $4 \times 4$ grid of dual-polarisation dipoles. Tile beams are formed by combining the dipole signals in an analog beamformer, using a set of switchable delay lines to provide coarse pointing capability \citep{Tingay:2013}. The beamformer cannot provide continuous tracking, instead it uses a combination of discrete pointings and drift scans in what is referred to as ``drift and shift''. The small size of each tile results in a field-of-view of $\sim$$600$ square degrees at 150\,MHz. In normal MWA operations, full polarisation correlation products (XX, YY, XY, YX) are stored by default. The work presented here uses the MWA array configuration present in the first three years of operations \citep{Tingay:2013}.

\begin{table*}[t]
\centering
\begin{tabular}{l c c c c c c c}
\hline\hline
Band & $\nu_{\text{min}}$ & $\nu_{\text{max}}$ & $\delta\phi$ & max. scale & $\lvert\phi_{max}\rvert$ & $\sigma_{\nu}$ \\ [0.5ex]
(MHz)      & (MHz) & (MHz) & (rad\,m$^{-2}$) & (rad\,m$^{-2}$) & (rad\,m$^{-2}$) & (mJy\,PSF$^{-1}$\,RMSF$^{-1}$) \\ [0.5ex]
\hline
89 & $72.30$ & $103.04$  & 0.40 & 0.37 & 91.1 & 24 \\
118 & $103.04$ & $133.76$ & 1.0 & 0.63 & 263.7 & 14 \\
154 & $138.88$ & $169.60$ & 2.3 & 1.0 & 645.6 & 7 \\
185 & $169.60$ & $200.32$ & 3.9 & 1.4  & 1175.6 & 5 \\
216 & $200.32$ & $231.04$ & 6.2 & 1.9  & 1937.0 & 5 \\ [1ex]
\hline\hline
\end{tabular}
\caption{Observing specifications for each MWA band assuming 40\,kHz channel bandwidth. The lowest and highest observing frequency are specified by $\nu_{\text{min}}$ and $\nu_{\text{max}}$, respectively. $\delta\phi$ and $\lvert\phi_{max}\rvert$ are the resolution and Faraday depth range available in each band. Where the maximum scale is smaller then $\delta\phi$, Faraday thick structures cannot be resolved. $\sigma_{\nu}$ is the typical noise in Faraday depth cubes for a naturally-weighted 2-minute snapshot.}
\label{table:polpar}
\end{table*}

The MWA observing bands are listed in Table \ref{table:polpar}. These bands are contiguous apart from a gap between 133.8\,MHz and 138.9\,MHz to avoid interference from the ORBCOMM constellation of satellites that transmit in this frequency range \citep{Offringa:2015}. While the MWA is capable of operating at frequencies above $230$\,MHz, RFI and the degraded response of the formed beam makes such observations problematic.

Studies of linear polarisation at low frequencies are complicated by the effects of Faraday rotation, as the degree of rotation is proportional to the wavelength squared. This can lead to bandwidth depolarisation if channel bandwidths are not small enough. However, with decreased channel bandwidth, sensitivity per channel is limited. Rotation measure (RM) synthesis \citep{Brentjens:2005v441p1217} is typically adopted to determine the degree of Faraday rotation across the entire observed bandwidth. This takes advantage of the Fourier relationship between the complex polarized intensity as a function of wavelength squared and the Faraday dispersion function (FDF) $\bm{F}(\phi)$ which is the polarized intensity as a function of Faraday depth \citep{Burn:1966}.

The available resolution in Faraday space ($\delta\phi$) increases with the longest wavelength; the maximum range of Faraday depths that can be probed ($\lvert\phi_{max}\rvert$) is also dependent on the longest wavelength and the channel bandwidth \citep[See equations 61--63,][]{Brentjens:2005v441p1217}. Table \ref{table:polpar} summarises these parameters for each MWA frequency band and also lists the maximum scale size. When the maximum scale is smaller than the resolution it is not possible to resolve Faraday-thick structures. For the MWA, when using a single observing band, this is always the case. The channel width listed in Table \ref{table:polpar} is 40\,kHz; however, with the MWA correlator \citep{Ord:2015} this can be reduced to as low as 10\,kHz (in which case the listed maximum Faraday depth increases by a factor of four). In general, a 40 kHz channel width is adequate for most Galactic and extragalactic studies. However, finer channels may be warranted for studies of high-RM pulsars; e.g., observing PSR J0742$-$2822 with a RM of $\sim$150\,rad\,m$^{-2}$ (Table \ref{table:polsources}) would require 20\,kHz channel widths in the 89\,MHz band to avoid bandwidth depolarisation.

\begin{figure}[t]
\includegraphics[width=\linewidth]{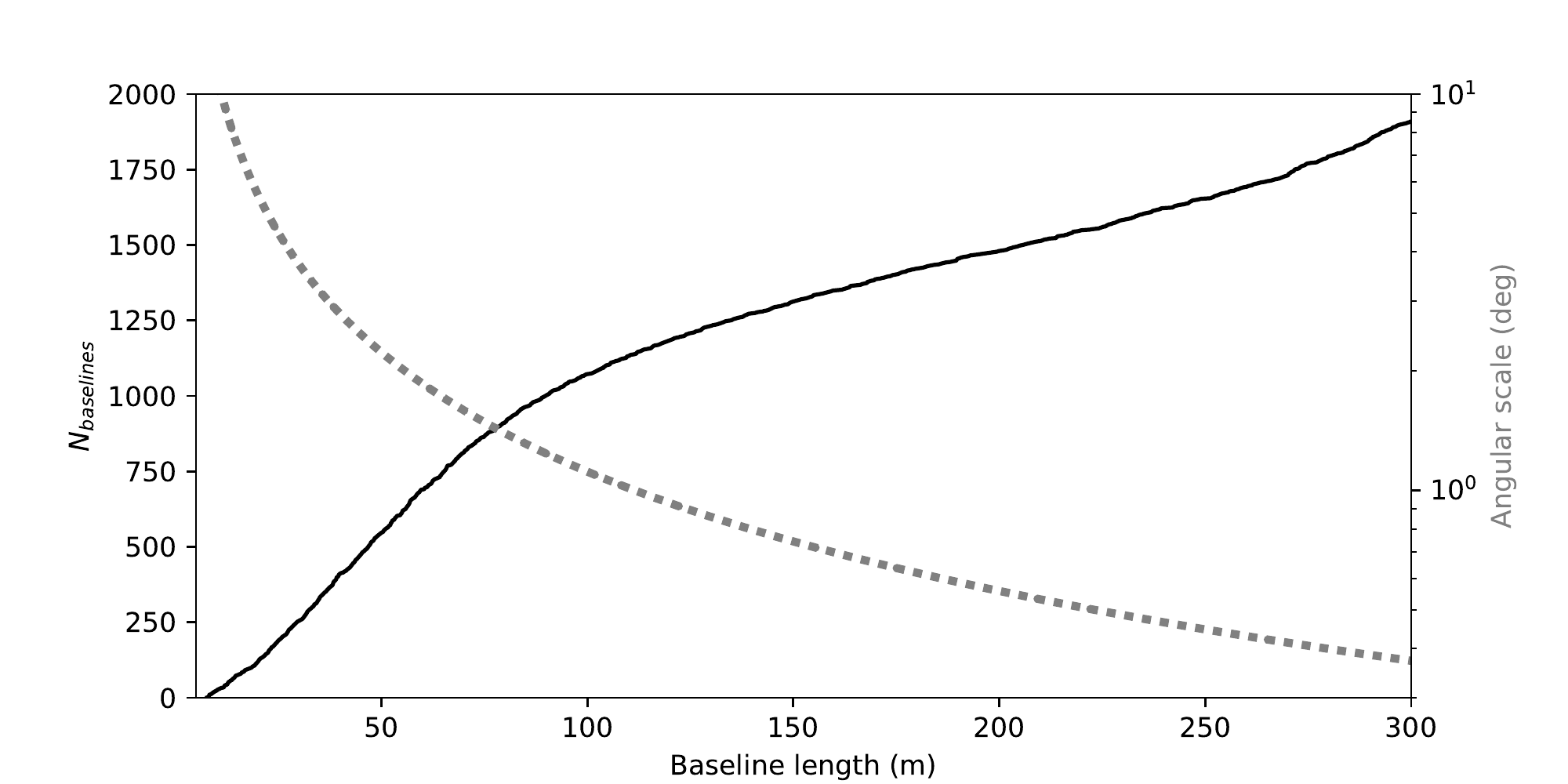}
\caption{The number of MWA baselines shorter than a given length (solid line) for a zenith pointing. The dotted grey line indicates the spatial-scale that a given baseline is sensitive to.}
\label{fig:bldist}
\end{figure}

The MWA has a compact and dense baseline configuration, with baselines as short as 7.7\,m. The availability of short baselines for large-scale structure studies is shown in Figure \ref{fig:bldist}; approximately 15\% (1183) of the baselines are shorter than 120\,m and provide sensitivity to structures a degree or more in extent. The MWA also provides baselines out to just under 3.0\,km in extent, almost an order-of-magnitude improvement in resolution compared to the 32T prototype \citep{Bernardi:2013v771p105}, yet still within the regime where the effect of the ionosphere on each tile can be assumed to be the same \citep{Lonsdale:2009,Arora:2016}.

The large number of baselines (8128) provides excellent $(u,v)$-coverage for snapshot imaging. This allows short time-scale imaging without any significant degradation in image quality. This feature is particularly useful for calibration testing and primary beam characterisation.

\subsection{Compact source polarimetry}
\label{sec:compact}

An early concern regarding polarimetry with the MWA was that beam depolarisation would effectively depolarise all intrinsically linearly polarised signals. Beam depolarisation occurs when there are spatial fluctuations in polarisation angle across the synthesised beam (which is of order $2\arcmin-5\arcmin$ with the MWA), resulting in a reduction in the measured linear polarisation. The 32-tile MWA prototype (32T) provided the first hints of what may be achievable with the full MWA. With baselines extending out to $\sim$$350$\,m, 32T provided limited $(u,v)$-coverage out to $\sim$$4\arcdeg$ scales and a resolution of $\sim$$16\arcmin$ at 189\,MHz. Despite this, 32T was able to detect a linearly polarised extragalactic source (PMN J0351$-$2744) and large-scale linearly polarised diffuse emission \citep{Bernardi:2013v771p105}.

\begin{figure}[ht]
\includegraphics[width=\linewidth]{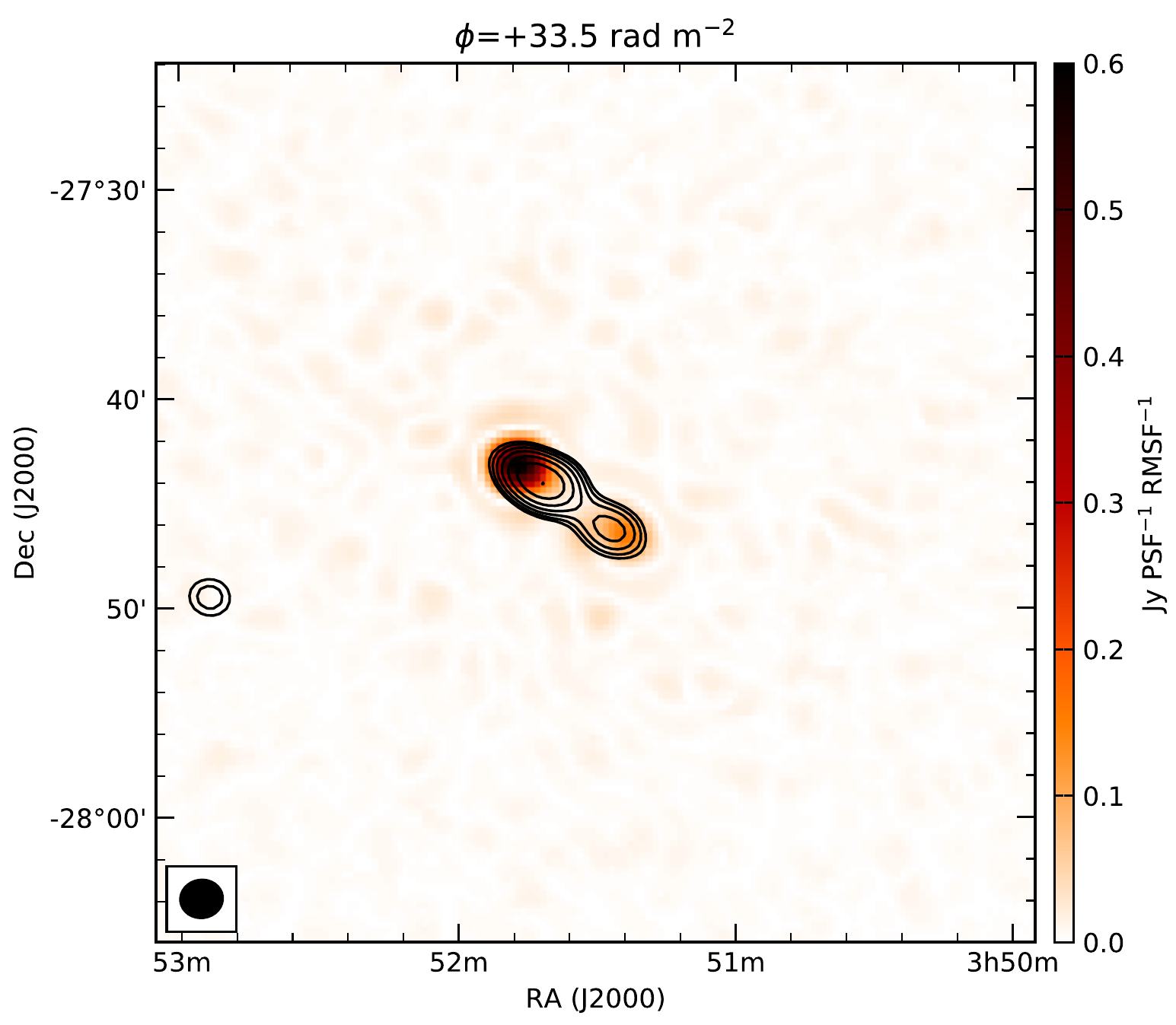}
\caption{Polarised intensity map of the source PMN J0351$-$2744 taken at a Faraday depth of $\phi$=+33.5\,rad\,m$^{-2}$ (corrected for ionspheric Faraday rotation, see Section \ref{sec:ionosphere} for details). Contours show total intensity with the lowest contour at 1.45 Jy\,PSF$^{-1}$ and subsequent contours increasing by a factor of $\sqrt{2}$. The synthesised beam size is $2\arcmin\times1.8\arcmin$ FWHM at a position angle of $-83\arcdeg$.}
\label{fig:j0351_128t}
\end{figure}

\begin{figure}[ht]
\includegraphics[width=\linewidth]{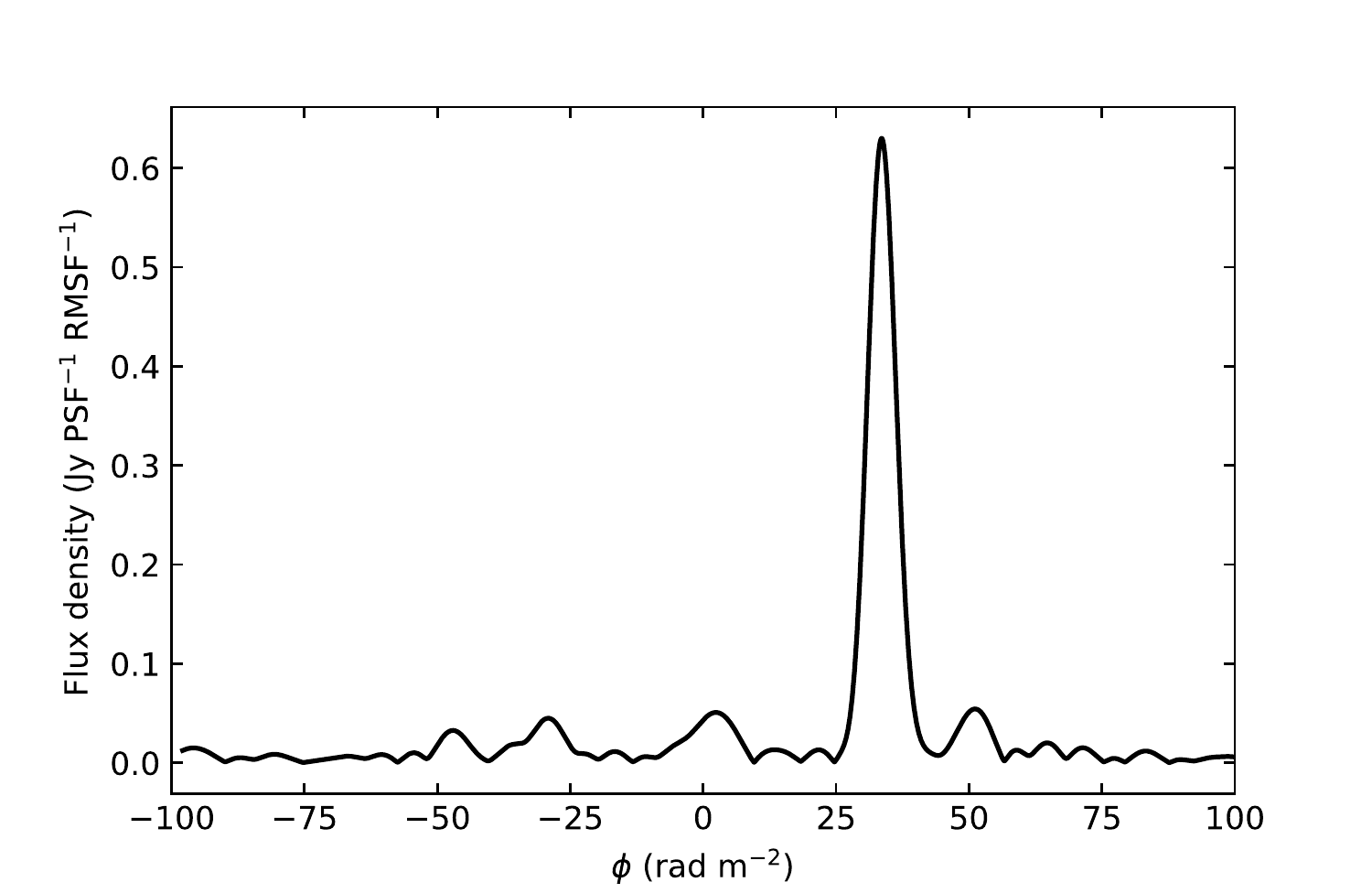}
\caption{Faraday dispersion function for the eastern hot spot of the polarised source PMN J0351$-$2744 at 216 MHz. The eastern hot spot peaks at a Faraday depth of $+33.6$\,rad\,m$^{-2}$ and a flux density of 630 mJy\,PSF$^{-1}$\,RMSF$^{-1}$. }
\label{fig:j0351_fdf}
\end{figure}

Observations from the MWA show PMN J0351$-$2744 is clearly resolved into a double linearly polarised source. The resolution of $\sim$$2\arcmin$ at 216\,MHz shows that the polarised component was associated with the edge of the source seen in total intensity (Figure \ref{fig:j0351_128t}). This shows that the scale size of the emitting regions in total intensity is generally greater than that for linear polarisation, so that the intrinsic fractional polarisation of the latter will be underestimated in observations with low spatial resolution. Figure \ref{fig:j0351_fdf} shows the FDF for the eastern hotspot. The measured RM of $+33.6$\,rad\,m$^{-2}$ for the hotspot is consistent with the RM determined by \citet{Bernardi:2013v771p105}.

We conducted a search for linearly polarised point sources using data taken as part of the GLEAM survey \citep{Wayth:2015v32p25,Hurley-Walker:2017A}. We limited the search to regions covered by the Murchison Widefield Array Commissioning Survey \citep{Hurley-Walker:2014v31p45} as the catalogue from this survey provided an excellent low frequency sky model for calibration purposes. Rotation measure synthesis and a conservative 10-sigma threshold was used to identify linearly polarised sources in the RM cube for each 2-minute snapshot. The search was shallow and only used data from the 216\,MHz band (to minimise frequency-dependent depolarisation effects), but it provided the first data to study the polarimetric characteristics of the instrument. Firstly, the beam model could be verified as sources transit through the beam. Secondly, polarised source candidates could be validated if they appeared in neighbouring 2-minute snapshots and with a similar RM and position. Thirdly, the flux, position and RM stability could be tracked for confirmed polarised sources. Finally, deeper images could then be made by integrating multiple snapshots in time.

We also made targeted measurements for 4348 known polarised sources listed in \citet{Taylor:2009v702p1230} and 72 known pulsars in \citet{Manchester:2005v129p1993}. In total, six sources were detected in a 6000 square-degree region, four were found in the $-27\arcdeg$ drift scan and two in the $-47\arcdeg$ drift scan. The sources and their polarisation parameters are summarised in Table \ref{table:polsources}.

\begin{table*}[t]
\centering
\begin{tabular}{l l l l l l}
\hline \hline
Source name & Type & P$_{\text{MWA}}$ & RM$_{\text{MWA}}$ & RM$_{\text{lit}}$ & Reference \\ [0.5ex]
 & & (mJy) & (rad\,m$^{-2}$) & (rad\,m$^{-2}$) \\
\hline
PMN J0351$-$2744 (E) & AGN Hot spot & 637 & $+$33.58$\pm0.03$ & $+$34.7$\pm5.5$ &  \citet{Taylor:2009v702p1230} \\
PMN J0351$-$2744 (W) & AGN Hot spot & 148 & $+$34.9$\pm0.2$ &  &   \\
PSR J0437$-$4715 & Pulsar & 195 & $+$2.2$\pm0.1$ & $+$0.58$\pm0.09$ &  \citet{Dai:2015} \\
PSR J0630$-$2834  & Pulsar & 151 & $+$46.5$\pm0.1$ & $+$46.53$\pm0.12$ &  \citet{Johnston:2005} \\
PKS J0636$-$2036 (N) & AGN Hot spot & 142 & $+$36.1$\pm0.3$ & $+$34.8$\pm7.2$ &  \citet{Taylor:2009v702p1230}  \\
PKS J0636$-$2036 (S) & AGN Hot spot & 1283 & $+$50.18$\pm0.05$ & $+$47.1$\pm1.9$ &  \citet{Taylor:2009v702p1230}  \\
PSR J0742$-$2822 & Pulsar & 293 & $+$150.6$\pm0.1$ & $+$149.95$\pm0.05$ &  \citet{Johnston:2005} \\
PSR J0835$-$4510 & Pulsar & 2234 & $+$37.3$\pm0.1$ & $+$31.38$\pm0.01$ &  \citet{Johnston:2005} \\ [1ex]
\hline
\end{tabular}
\caption{Polarised point sources observed in the 216\,MHz band in a GLEAM zenith drift scan (2013 Nov 25 14:44 to 19:53) and a $-47\arcdeg$ meridian drift scan (2013 Nov 6 13:27 to 21:08). P$_{\text{MWA}}$ is the polarized flux density measured from MWA observations. RM$_{\text{MWA}}$ is the rotation measure determined from MWA observations (corrected for ionospheric Faraday rotation, see Section \ref{sec:ionosphere} for details). RM$_{\text{lit}}$ is the rotation measure in literature. The lowest frequencies used in the literature value RM measurements are 1340\,MHz \citep{Taylor:2009v702p1230}, $700$\,MHz \citep{Dai:2015}, and 1240\,MHz \citep{Johnston:2005}. The polarised eastern/western components for PMN J0351$-$2744 and northern/southern components for PKS J0636$-$2036 are listed separately.}
\label{table:polsources}
\end{table*}

Four of the six sources detected are pulsars and are the four brightest known pulsars listed in the ATNF Pulsar Catalogue\footnote{\url{http://www.atnf.csiro.au/research/pulsar/psrcat}} v1.54 \citep{Manchester:2005v129p1993} in the 6000 square-degree region surveyed (a source density of one per 1500 square-degrees, see Section \ref{sec:mwapoint} for further discussion). Pulsars are easier to detect in polarisation compared to other polarised sources at MWA wavelengths because (a) they generally have high fractional polarisation (so are easier to separate from polarisation leakage); and (b) they have negligible angular extent (and so don't suffer from depolarisation). This may make polarisation searches an interesting technique for discovering pulsars in imaging surveys; see Section \ref{sec:pulsars}.

All of the RMs measured for the polarised sources are consistent with RM values found in literature (with an additional $\pm2$ rad m$^{-2}$ margin of error where ionospheric corrections were not made in the literature) apart from PSR J0835$-$4510. This is the Vela pulsar and has an RM that varies on a time-scale of about a year as a result of being embedded in an ionised cloud formed by a pulsar wind nebula and the Vela supernova remnant \citep{Hamilton:1977,HamiltonB:1985,Johnston:2005}. 

PKS J0636$-$2036 is the brightest linearly polarised extragalactic source detected with the MWA to date. It is also the only extragalactic source that has been detected in all five of the MWA frequency bands. As a result, it has become the default source for testing polarimetric behaviour of the MWA. 

\subsection{Circular polarisation}
\label{sec:circpol}

A majority of astronomical sources do not emit significant circular polarisation. As a result, images of circular polarisation (Stokes V) are not limited by source confusion or sidelobe confusion and can achieve sensitivity levels that approach thermal noise. This also means that the angular resolution of the instrument is not critical for detection and is only required to improve the localisation of the source.

Circular polarisation is not affected by Faraday rotation and can be processed in a similar manner to Stokes I imaging by treating the entire frequency band in a continuum-like mode (assuming the sign of circular polarisation does not change across the band). However, as many sources which emit in Stokes V are particularly energetic and may exhibit flaring or transient behaviour, care must be taken in choosing the time-scale over which integration is performed when searching for variable circular polarisation. If the event is brief but integrated over an extended period of time then the signal associated with the event will be diluted. Similarly, if the sign of polarisation flips over the course of the integration \citep[e.g.][]{Lynch:2017a, Lynch:2017b} then this can reduce the signal.

The first sources detected in circular polarisation with the MWA were pulsars. For example PSR J0835-4510 exhibits circular polarisation across all five MWA frequency bands and PSR J0437-4715 is also clearly detected in Stokes V images. \citet{Lenc:2016} demonstrated that long integrations in Stokes V could be used to detect weak pulsars in deep fields. This technique could be used to detect pulsars and other circularly polarised sources such as flare stars \citep{Lynch:2017b} and potentially extrasolar planets \citep{Murphy:2015v446p2560}.

\subsection{Diffuse polarisation}
\label{sec:diffuse}
Low frequency observations of diffuse polarisation are sensitive to small changes in Faraday rotation as a result of fluctuations in the magnetized plasma, which are difficult to detect with observations at high frequencies. Early low frequency observations with synthesis telescopes e.g. WSRT between $325$ and $375$\,MHz \citep{Wieringa:1993,Haverkorn:2000,Haverkorn:2003A,Haverkorn:2003B,Haverkorn:2003C} and WSRT at 150\,MHz \citep{Bernardi:2009, Bernardi:2010} were only sensitive to structures smaller than $\sim$$1\arcdeg$. The structures detected were weak and would require significant integration time to detect with the MWA.

\begin{figure*}[t]
\includegraphics[width=\linewidth]{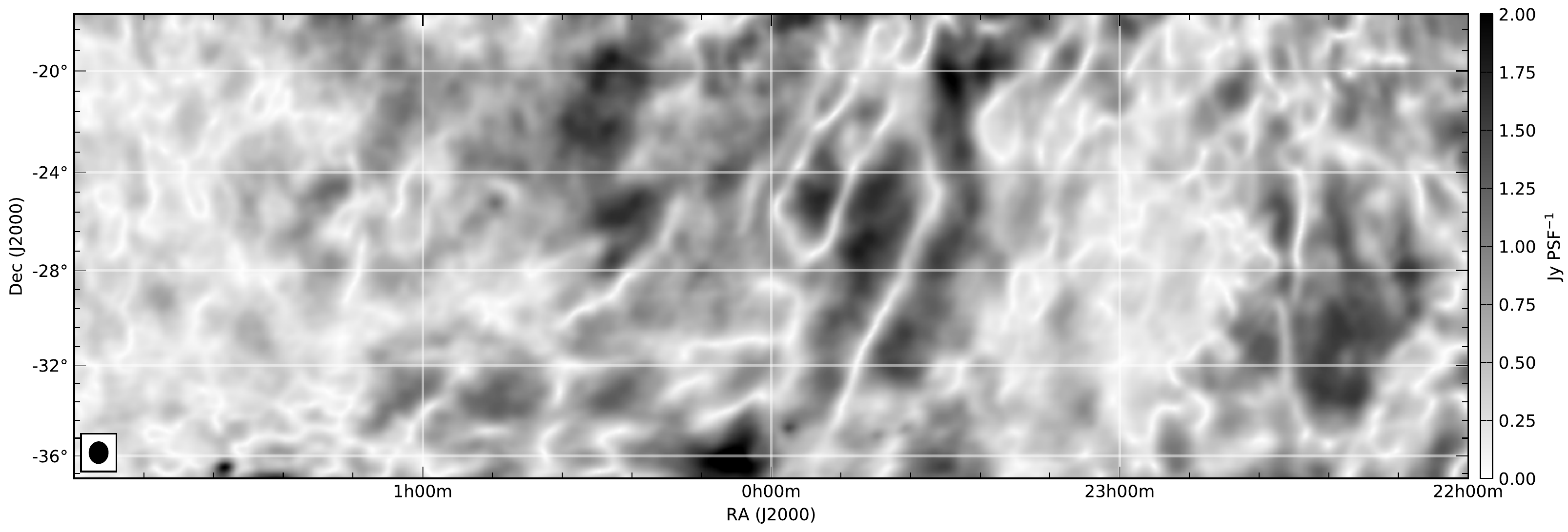}
\caption{Polarised intensity observed over a zenith drift scan (not corrected for ionospheric Faraday rotation) in the 216 MHz band. The flux scale is in Jy\,PSF$^{-1}$. The synthesised beam, shown as a filled ellipse, is 54$\arcmin\times$47$\arcmin$ FWHM at a position angle of -1.8$\arcdeg$.}
\label{fig:diffuse}
\end{figure*}

More recent observations with LOFAR at 150\,MHz \citep{Iacobelli:2013,Jelic:2015,VanEck:2017} achieved sensitivity to spatial scales up to $\sim$$5\arcdeg$ by utilising a high band antenna dual-inner mode \citep{vanHaarlem:2013}. While these observations also provided baselines significantly longer than currently available with the MWA, enabling a study of weak diffuse structures at a resolution hardly affected by beam depolarisation \citep{Jelic:2015}, they provided relatively few short baselines and so sensitivity to large scale structures was limited.  Single dish polarimetric observations at long wavelengths provide access to large-scale structure, e.g. \citet{Mathewson:1965}, but single dish observations below 300\,MHz lack resolution and sensitivity.

Observations at 189\,MHz with the MWA 32T \citep{Bernardi:2013v771p105} provided the first indication of what the MWA may be able to detect. Its relatively compact baselines provided limited sensitivity out to $\sim$$4\arcdeg$ scales, which was enough to demonstrate that linearly polarised diffuse emission from the local interstellar medium could be detected with an MWA-like array. 

When the MWA was used to observe diffuse linear polarisation, a surprising outcome was that the emission was bright and present on larger scales than previously observed \citep{Lenc:2016}. Figure \ref{fig:diffuse} shows an example of the diffuse emission observed at 216 MHz with the MWA; the image was formed by combining $24\times2$ minute snapshots from a zenith drift scan. The polarised emission is so dominant that it is easily detected in 2-minute snapshot images with a signal-to-noise of $\sim$$70$.

The high signal-to-noise at which the MWA detects this diffuse polarised structure provides a valuable resource. For example, XY-phase calibration ideally requires the presence of a bright polarised source (see Section \ref{sec:xyphase}). Similarly, the emission can be used to monitor the polarimetric response across the beam and can provide a reference against which temporal changes in Faraday rotation can be monitored.

\section{Processing for polarimetry}
\label{sec:mwasw}

Most radio data reduction packages were not designed with fixed dipole-based instruments in mind. As such, they have built in assumptions that are generally only valid for targeted, dish-based instruments. Nonetheless, packages such as \textsc{Miriad} \citep{Sault:1995} and \textsc{casa} \citep{McMullin:2007} were used in early MWA processing pipelines \citep{Hurley-Walker:2014v31p45} and are suitable for basic calibration and imaging in Stokes I. 

However, these packages are unable to solve for full wide-field polarimetric effects and so their use for polarimetry is restricted to narrow-field imaging of Stokes Q at zenith. Generating the full spectral cubes required for polarimetry is computationally expensive as these packages were not designed to work within modern high-performance computing (HPC) environments.

Recent processing pipelines for the MWA use packages that are designed to work in HPC environments and the specific calibration and imaging needs of fixed dipole-based instruments. The three packages in use for polarimetry are: the Real Time System \citep[\textsc{rts},][]{Mitchell:2008v2p707,Ord:2010}; \textsc{mitchcal} \citep{Offringa:2016v458p1057} together with \textsc{wsclean}\footnote{\textsc{wsclean}: \url{ http://sourceforge.net/p/wsclean}} \citep{Offringa:2014v444p606O}; and Fast Holographic Deconvolution\footnote{\textsc{fhd}: \url{https://github.com/EoRImaging/FHD}} \citep[\textsc{fhd},][]{Sullivan:2012v759p17}. These packages are described in further detail in the following sections. Note that the \textsc{rts} package was used for all processing described in this paper.

It is worth briefly mentioning the typical visibility weighting schemes used for MWA polarimetry. For polarimetry of diffuse polarisation, natural weighting works well with the dense baseline distribution in the MWA core. This can be further optimised by removing baselines outside of the MWA core as it results in a more Gaussian synthesised beam \citep[see][]{Lenc:2016}. For compact source polarimetry, robust weighting (with robustness of $-$1) works well. Further improvement is obtained by removing the shortest baselines from the MWA core to avoid excessive down-weighting of the longest baselines.

\subsection{Real Time System}

The \textsc{rts} is the only one of these packages that can take advantage of Graphics Processing Units (GPUs) to increase performance (it can also run on CPUs if GPUs are unavailable).  The \textsc{rts} calibrates based on a local sky model and can generate full spectral resolution Stokes cubes that are corrected for dipole projection effects and wide-field effects across the entire field of view. These can be produced in real-time for small fields of view with limited peeling or near real-time for larger fields.

Prior to imaging, the \textsc{rts} can optionally peel out sources in a local sky model from visibility data to reduce Point Spread Function (PSF) sidelobe confusion. It also solves for ionospheric positional shifts and flux density variations during the peeling process. 

One drawback of the \textsc{rts} is that it cannot perform deconvolution. This is not critical for polarimetry but it can help to reduce PSF sidelobes of particularly bright Stokes I sources that leak signal into the polarisation products. If deconvolution is necessary, the \textsc{rts} can export calibrated and peeled visibilities in FITS format and these can be deconvolved outside of the package \citep[preferably with an algorithm that deals appropriately with the complex nature of polarisation, see][]{Pratley:2016}. Another constraint is that the imaging field of view in GPU mode is limited by GPU memory constraints (this is less of an issue in CPU mode).

\subsection{MitchCal and wsclean}

\textsc{MitchCal} \citep{Offringa:2016v458p1057} is a custom implementation of the direction-independent full-polarisation self-calibration algorithm used by the \textsc{rts} \citep{Mitchell:2008v2p707}. In combination with \textsc{wsclean}, this set of tools can generate full Stokes images. The tools have mostly been used to generate and deconvolve continuum maps but can be adapted to generate spectral cubes.

As with the \textsc{rts}, this set of tools provides a means to calibrate for ionospheric shifts. It also provides direction dependent calibration by clustering sources to improve calibration signal-to-noise in the direction of the cluster.  A further advantage, in comparison to the \textsc{rts}, is the ability to perform efficient wide-field deconvolution.

\subsection{FHD}

Fast Holographic Deconvolution \citep[\textsc{fhd},][]{Sullivan:2012v759p17} is an imaging and calibration algorithm designed for wide field-of-view interferometers with direction- and antenna-dependent beam patterns. It has been used for EoR science \citep{Jacobs:2016}, but it includes functionality for calibration, imaging, deconvolution and simulation that can be applied more generally. 

Using a widefield polarized antenna beam model, FHD calibrates the X and Y dipoles separately to match an input sky model (typically $10^{4}$ sources). The global phase between the X and Y dipoles is fit by minimizing the apparent V power, explicitly assuming there is very little real V power on the sky and little of the Q polarization is mixed to V in a crossed dipole array. If given a sky model with polarized emission, FHD can fit a global ionospheric Faraday mixing angle between Q and U. This mixing angle is used to correct for ionospheric rotation when making the diffuse image cubes.    

Deconvolution in full Stokes I, Q, U, and V is possible with FHD. This can be used to generate models in other Stokes parameters besides I, and may be essential to making diffuse maps without bright Stokes I leakage. FHD also has the ability to generate a separate beam model for each antenna to help reduce leakage if that level of detail is known.

\section{MWA polarisation calibration}
\label{sec:mwacal}

There are three aspects of MWA calibration that are particularly relevant to polarimetry: calibrating for the primary beam shape, calibrating for XY-phase and correcting for ionospheric Faraday rotation. We now discuss each aspect in turn.

\subsection{Beam model}
\label{sec:beam}

Early analysis of MWA drift scan images revealed that many sources appeared to have significant levels of linear and circular polarisation towards the edge of the primary beam. This effect is often referred to as ``leakage'', where emission from Stokes I appears to leak into the other Stokes parameters. For drift scans passing through zenith ($-$27 degrees) and observed in the 154\,MHz band, the level of leakage was consistent with that measured by \citet{Bernardi:2013v771p105}; typically of the order of $\sim$1\% in the beam centre and $\sim$4\% towards the edge of the field. 

\begin{figure*}[t]
\includegraphics[width=0.5\linewidth]{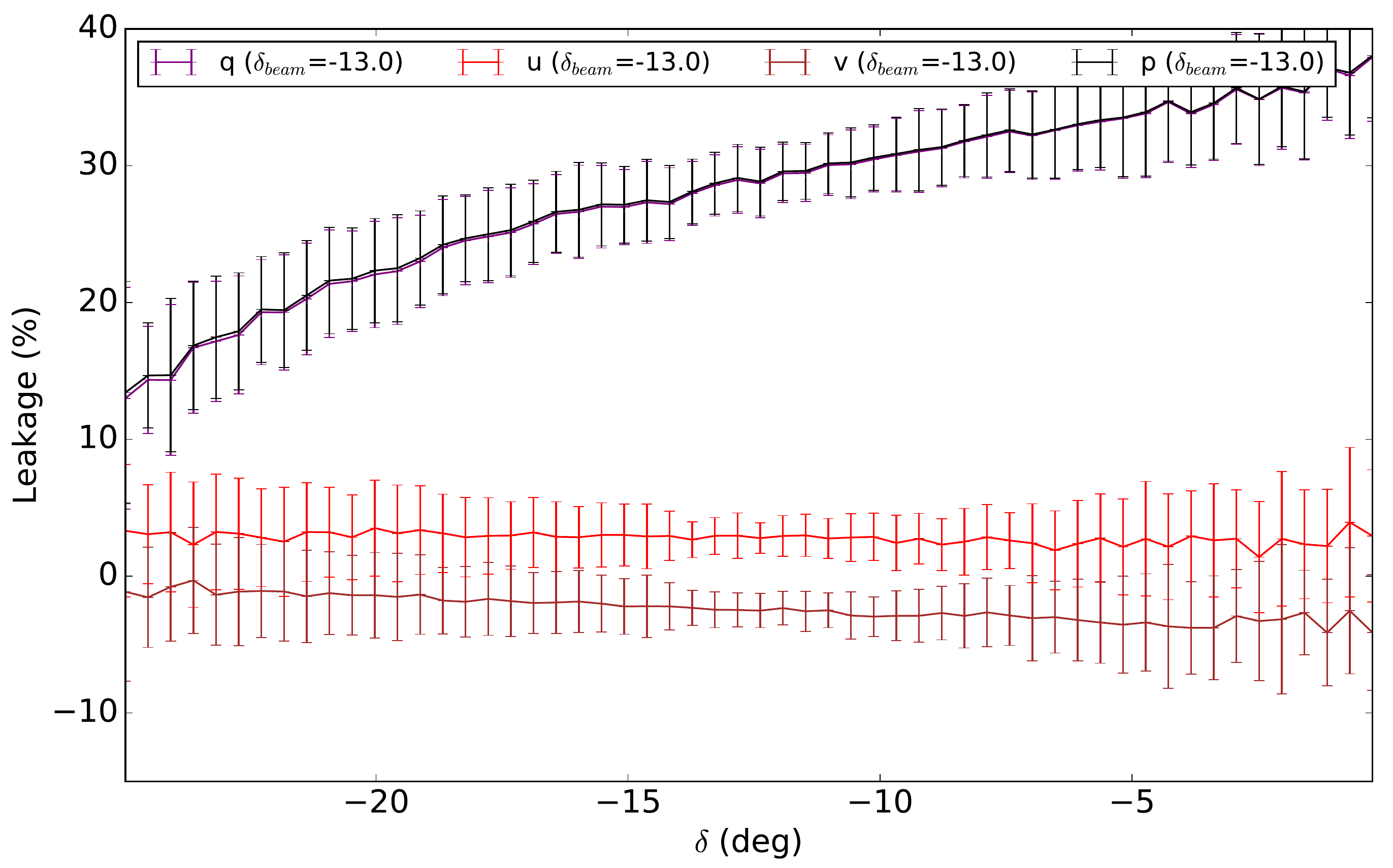}
\includegraphics[width=0.5\linewidth]{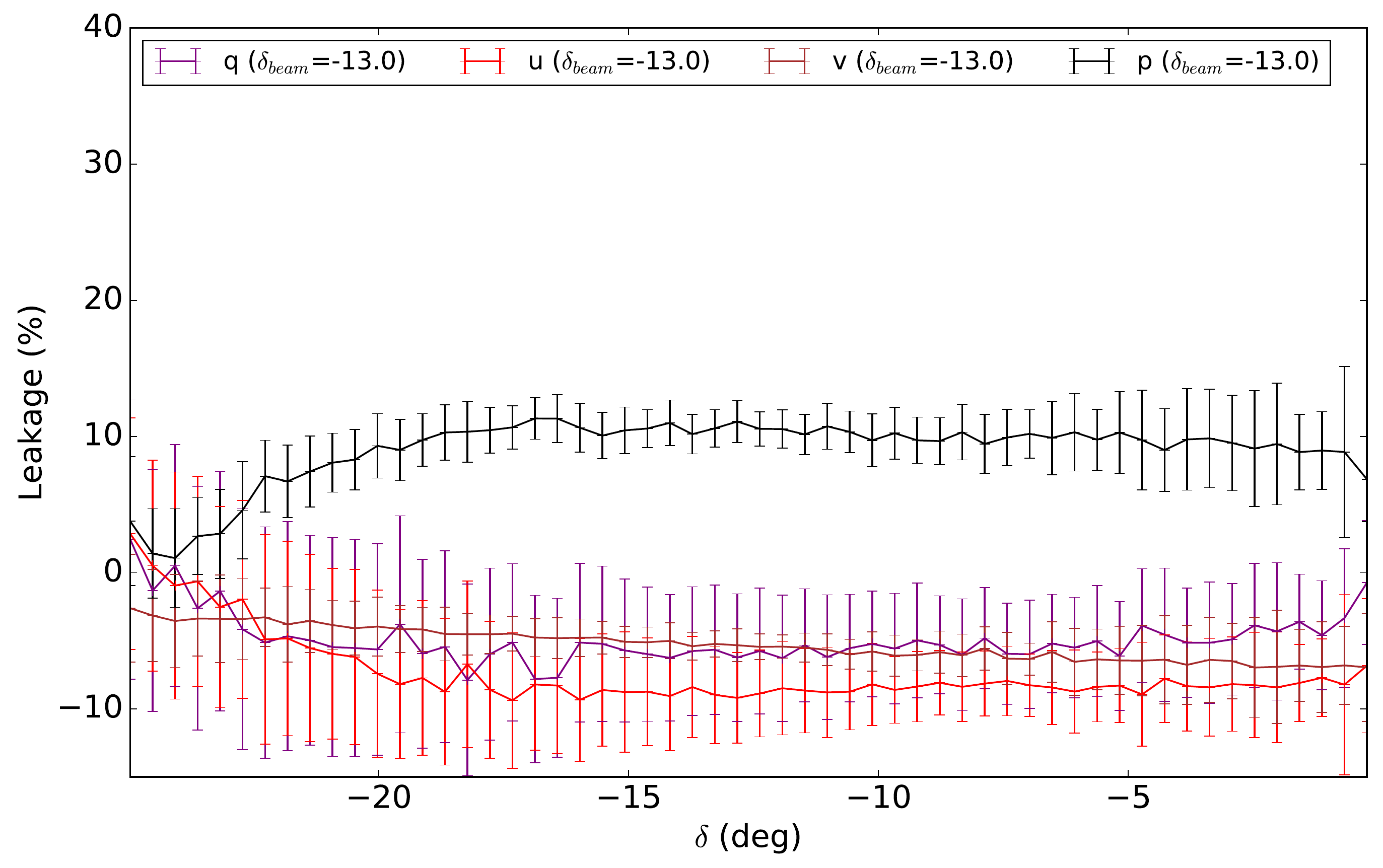}
\caption{Left: Declination-dependent leakage observed in the MWA 216\,MHz band for a $\delta=-13\arcdeg$ declination meridian scan using an analytic beam model to correct for instrumental polarisation. The plot shows the median fractional polarisation for Stokes Q, U and V where $q=Q/I$, $u=U/I$ and $v=V/I$. The fractional linearly polarised intensity $p$ is also shown where $p=\sqrt{Q^{2}+U^{2}}/I$. Right: Same as left but using the \citet{Sutinjo:2015v50p52S} beam model.}
\label{fig:beam}
\end{figure*}

However, for beamformer pointings away from zenith and for observations in the highest frequency bands the measured leakage was higher. For example, the plot on the left in Figure \ref{fig:beam} shows the measured percentage leakage in Stokes Q, U and V measured in the MWA 216-MHz band for field sources as a function of declination for a $-13\arcdeg$ declination meridian scan. In this plot, the leakage in Q increases from 12\% near zenith to nearly 40\% at $0\arcdeg$ declination.  While leakage in U and V also show declination dependent trends they occur at much lower levels and are not as problematic.

These observed trends are symptomatic of errors in the primary beam model. Stokes Q is generally affected the most as it is formed from the same instrumental polarisations as Stokes I, i.e. I$=\frac{1}{2}($XX$+$YY$)$ and Q$=\frac{1}{2}($XX$-$YY$)$, and so these contain the bulk of the total intensity signal. Stokes U and V are formed from the cross polarisations (XY and YX) which do not have significant signal content.  As such, any errors in correcting for the XX and YY for the beam shape using a primary beam model result in some portion of a large signal leaking into Stokes Q. Primary beam model errors can also result in a scaling error in Stokes I, as noted by \citet{Hurley-Walker:2014v31p45} and \citet{Hurley-Walker:2017A}. This error manifested itself as a declination and frequency dependent scaling error in Stokes I.

Early imaging efforts with the MWA used an analytic beam model of a simple short-dipole with an ideal array factor derived by \citet{Bowman:2007}. This relatively simple model was a good approximation of an MWA tile at an optimal observing frequency of 154 MHz and towards the zenith. However, the model diverges from observations at higher frequencies bands where the simple short-dipole model no longer holds. Two approaches were taken in an attempt to reduce such errors: improvements in the beam model, and mitigation techniques using empirical data. These are described in detail in the following sections.

\subsubsection{Beam improvements}
\label{sec:newbeam}

To improve the beam model, \citet{Sutinjo:2015v50p52S} performed simulations that incorporated both mutual coupling and embedded element patterns. Initial testing of an implementation of this model shows an improvement in overall polarisation leakage. The plot on the right in Figure \ref{fig:beam} shows the improvement obtained in in comparison to the analytic beam model. Leakage from Stokes I to Stokes Q shows a dramatic reduction in the new beam model, however, there is an associated increase in leakage measured in Stokes U and V.

Despite the slight increase in leakage in Stokes U and V, the overall factor of 3$-$4 reduction in polarisation leakage in this beamformer pointing and frequency band is a significant improvement. It should be noted that this a worst-case scenario as it highlights one of the most severely affected bands and beamformer pointings; furthermore, no attempt was made to limit the leakage with in-beam self-calibration. Further testing is currently being performed on the improved beam model to compare its performance against the analytic beam for various beamformer pointings and observing bands (Sokolowski, in prep.).

\subsubsection{Mitigating leakage in polarisation}
\label{sec:mitigateleakage}

Despite the improvements in primary beam models, there will always be some errors due to imperfections in the models and in the MWA dipoles themselves. Differences in orientation, environment and the number of dipole failures within a tile affect the true beam response of the tile. The more complex the beam model becomes, the more computationally intensive it is to calculate during the calibration and imaging process.

\begin{figure*}[t]
\includegraphics[width=\linewidth]{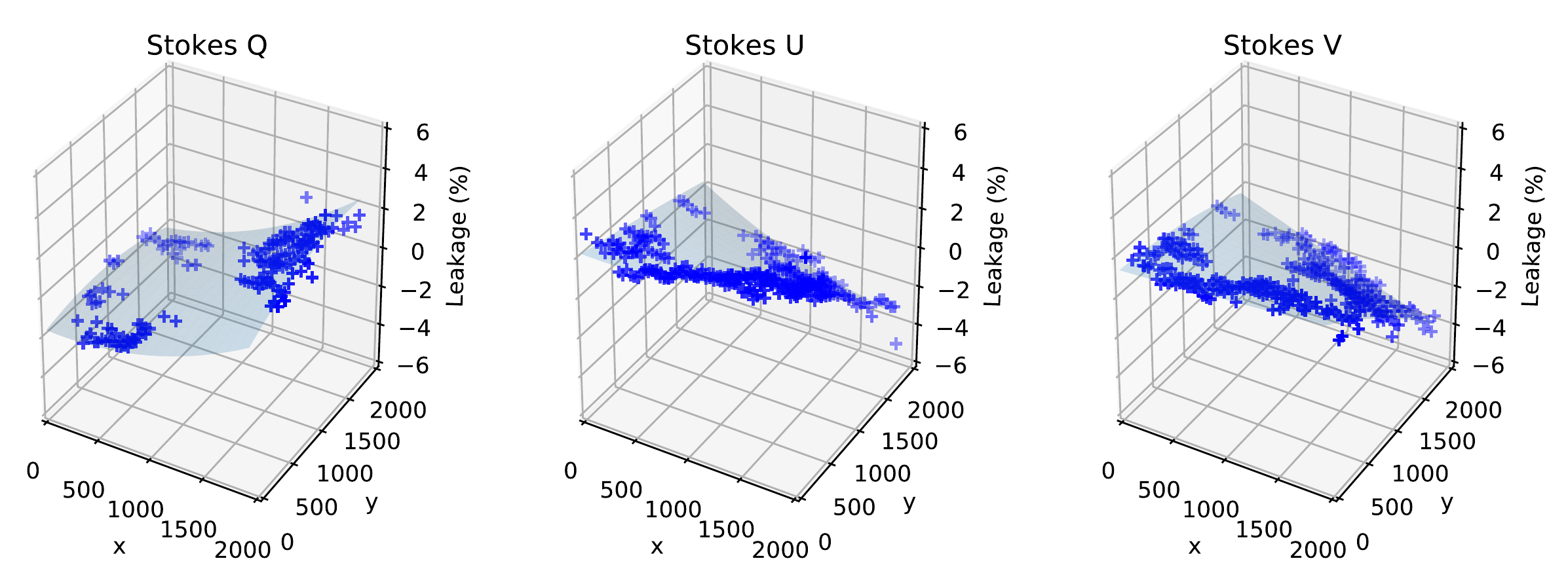}
\caption{A sample demonstrating the fitting of point-source leakage over the observed field-of-view. The points were taken from one of the beamformer pointings used during a ``drift and shift'' observation. Fits were performed separately for different beamformer pointings and for each of the different Stokes parameters (Q, U and V). The X and Y axis are in units of pixels for a $25\arcdeg\times25\arcdeg$ field.}
\label{fig:fitbeam}
\end{figure*}

For small fields of view around a bright source, in-beam calibration with a sky model can reduce beam-related leakage effects close to the dominant source of the sky model by assuming it is unpolarised. In general, this is a reasonable assumption given that few sources are polarised at long wavelengths and those that are typically have a fractional polarisation of a few percent. However, this only reduces leakage in the vicinity of the dominant source. Further away from the dominant source,
polarisation leakage once again becomes evident.

An alternative to in-beam calibration is measuring and removing the beam-related errors empirically. The field-of-view and $(u,v)$-coverage of the MWA means that the beam can be mapped with short (2-minute) snapshots over the course of an extended drift scan. This approach was taken by \citet{Hurley-Walker:2017A,Callingham:2017} to correct for declination and frequency-dependent flux-scale errors observed in Stokes I for point sources as they drifted through the MWA beam. These corrections also highlighted spatial calibration discrepancies in other long wavelength surveys \citep{Hurley-Walker:2017B}.

\begin{figure*}[t]
\includegraphics[width=\linewidth]{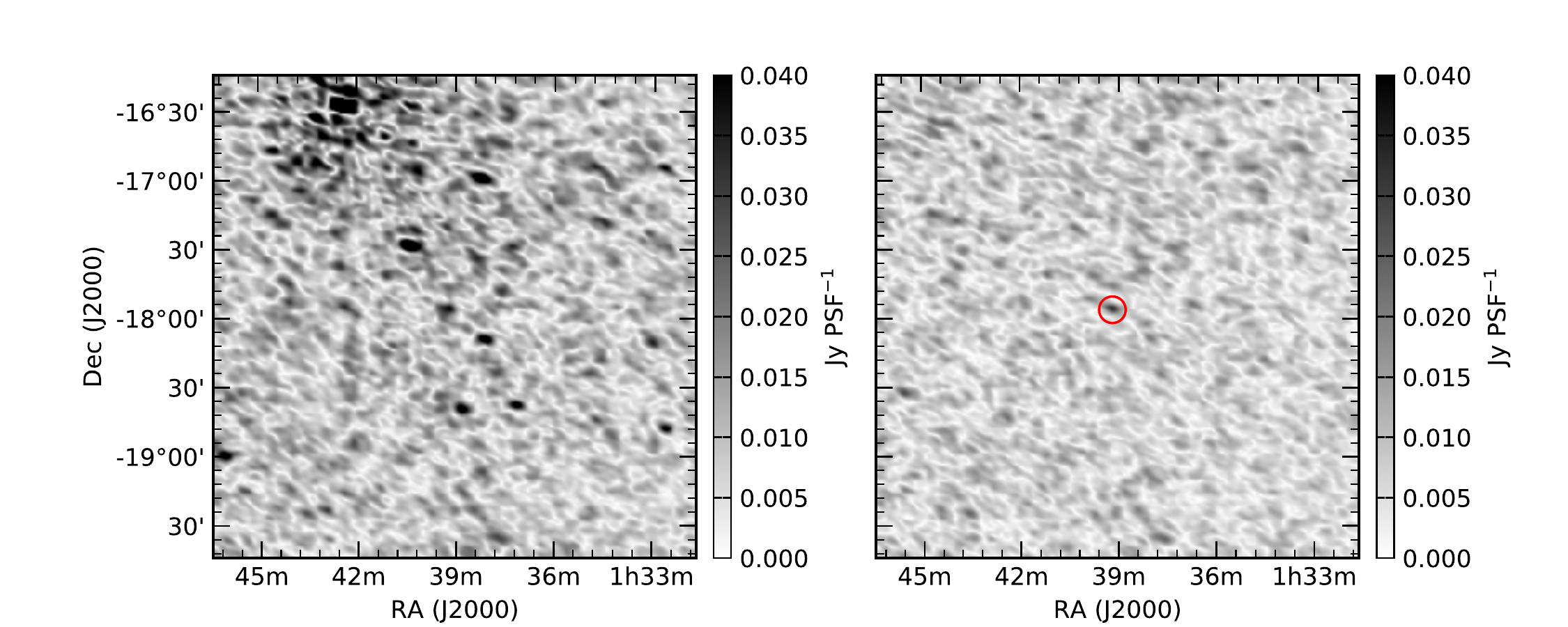}
\caption{The polarised intensity map for the UV~Ceti field before leakage subtraction in Stokes Q and Stokes U (left), and after leakage subtraction (right). In the leakage-subtracted image the dominant source in the field (highlighted with a red circle) is UV~Ceti shown during a flare.}
\label{fig:leaksub}
\end{figure*}

A similar approach can be applied to Stokes Q, U and V by mapping the observed leakage of point sources as they drift through a given beamformer pointing. By assuming sources are unpolarised, the leakage can be modelled spatially across the beam. Figure \ref{fig:fitbeam} shows the measured leakage in field sources for one beamformer pointing in observations of the flare star UV~Ceti.

A spatial fit of the leakage can be determined for each Stokes parameter and for each beamformer pointing. The fit is then used to scale and subtract the Stokes I images from each of the Stokes Q, U and V images, removing the leaked component of Stokes I. Figure \ref{fig:leaksub} demonstrates the dramatic improvement that results from this method and the subsequent detection of a linearly polarised flare in UV~Ceti \citep{Lynch:2017a}. Prior to subtraction the field was dominated by polarisation leakage but after subtraction only a single source, UV~Ceti, dominated the field with 10$\sigma$ significance.

\subsection{X-Y phase calibration}
\label{sec:xyphase}

When calibration of linearly polarised antennas is performed on an unpolarized source, the phase relationship between the orthogonal X and Y polarisations cannot be solved for. Any residual phase results in leakage from linear polarisation into circular polarisation (Stokes V). To solve for the X-Y phase, calibration against a polarised source is required \citep{Sault:1996}. 

\begin{figure*}[t]
\includegraphics[width=\linewidth]{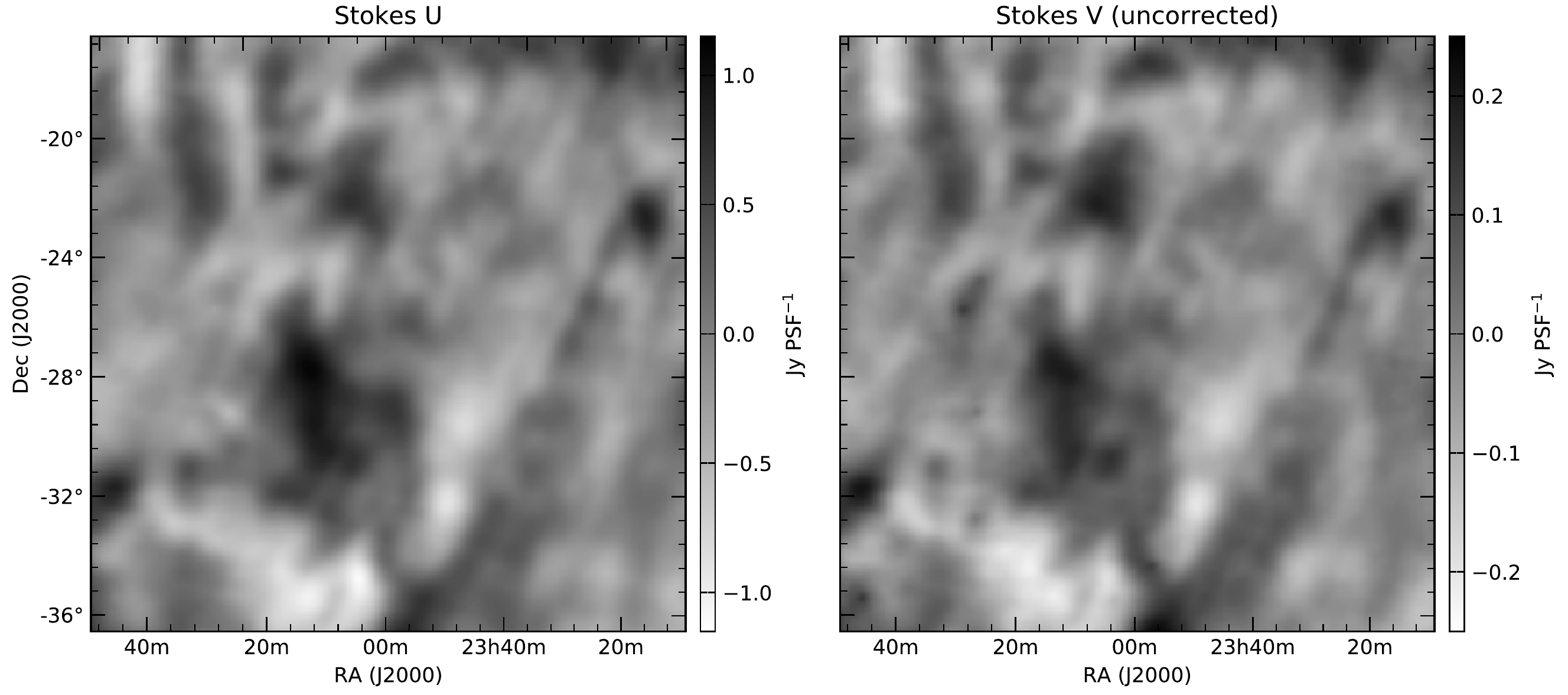}
\includegraphics[width=\linewidth]{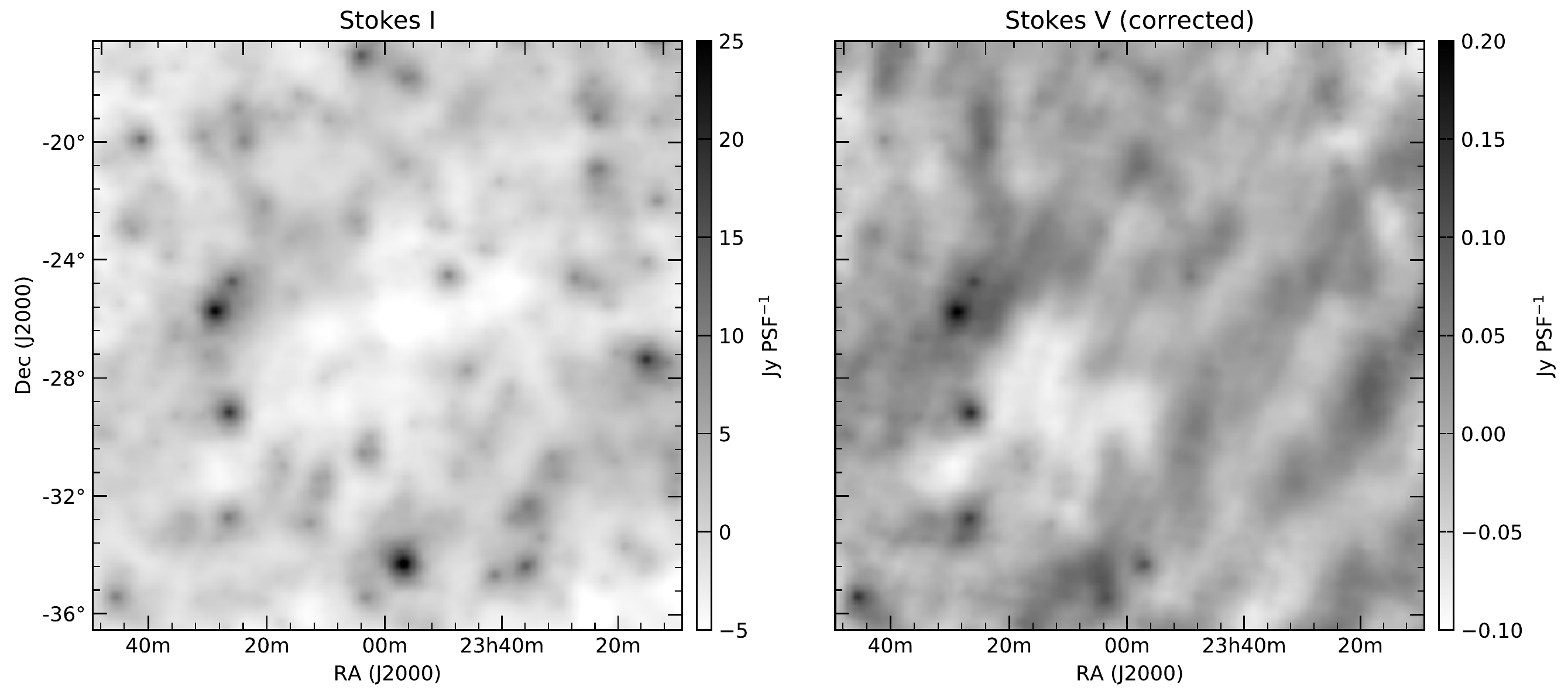}
\caption{The top two figures show the similarity between Stokes U (left) and Stokes V (right) in MWA observations of the EoR-0 field at 154 MHz. This  demonstrates the effect of leakage from Stokes U into Stokes V caused by uncorrected X-Y phase. The bottom two figures show Stokes I (left) and Stokes V (right) after applying a correction for the X-Y phase shown in Figure \ref{fig:xyphase}. The remaining features in Stokes V are dominated by leakage from Stokes I. This leakage is associated with errors in the beam model, see Section \ref{sec:beam}. Note the different flux density scales in the different images.}
\label{fig:uvleak}
\end{figure*}

Traditionally, X-Y phase is calibrated against a known polarisation point-source calibrator. The challenge is that compact linearly polarised sources suitable for calibration \citep{Bernardi:2013v771p105} are rare at long wavelengths (see Section \ref{sec:compact}).

An alternative is to use diffuse polarised emission. As shown in Section \ref{sec:diffuse}, the dense MWA core provides excellent sensitivity to large-scale structure. The diffuse polarised structure from the ISM has sufficient signal-to-noise to enable the X-Y phase to be solved for.

Figure \ref{fig:uvleak} shows the Stokes U and Stokes V images from observations at 154 MHz \citep{Lenc:2016}. No circular polarisation is expected in diffuse emission from the local ISM, yet there are clear diffuse structures in the Stokes V map. These structures closely resemble those that appear in Stokes U but at about the $20-30$\% level.

\begin{figure}[t]
\includegraphics[width=\linewidth]{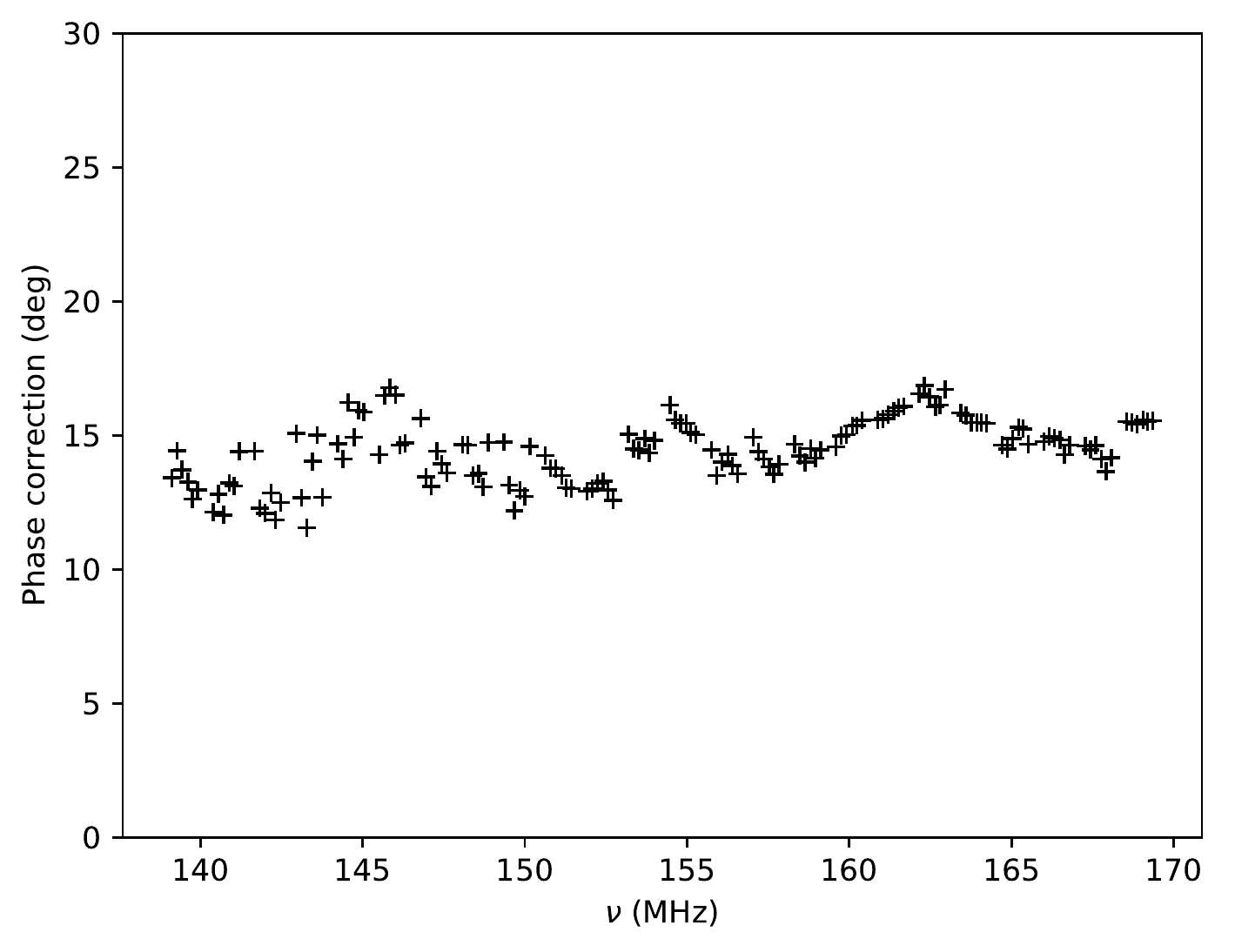}
\caption{X-Y phase measured as a function of frequency using diffuse polarised emission from the EoR-0 field.}
\label{fig:xyphase}
\end{figure}

Diffuse linear polarisation is evident across the southern sky, as shown in Figure \ref{fig:diffuse}. The ubiquity and strength of this signal makes it an ideal alternative for use in X-Y phase calibration. The emission appears with sufficient signal-to-noise to allow the X-Y phase to be measured as a function of frequency, see Figure \ref{fig:xyphase}, demonstrating that the phase is independent of frequency.

Applying a correction for the X-Y phase derived from the observed features in Stokes U and Stokes V results in a $5-10\%$ improvement in Stokes U and a $20-30$\% improvement in Stokes V. After the correction for X-Y phase, the corrected Stokes V image (Figure \ref{fig:uvleak}) is now dominated by leakage from Stokes I at the $\sim$$1\%$ level. This residual leakage results from errors in the beam model, as described in Section \ref{sec:beam}.

\subsection{Ionospheric Faraday rotation}
\label{sec:ionosphere}

Long-wavelength polarimetry allows us to measure Faraday rotation with high precision. So much so, that Faraday rotation introduced by the ionosphere is the dominant source of error in the observed rotation measure. In typical MWA nighttime observations at high elevations this effect is small. However, applications that require high-accuracy measurements of source RM still need to correct for the ionospheric component. Similarly, to avoid depolarisation, observations require calibration for ionospheric Faraday rotation when integrating over time-periods throughout which the ionospheric conditions vary significantly. The time period over which this can be critical can be on the order of an hour in the lowest MWA band (over which the Faraday rotation can vary on the order of $\sim$$0.3$ rad m$^{-2}$ at zenith during calm ionospheric conditions at night). The highest MWA band is more resilient to ionospheric effects, but corrections must still be considered if integrating data from separate days in which the ionospheric conditions varied significantly (such variations can typically be of order several rad m$^{-2}$).

\subsubsection{Ionospheric modelling}

A number of tools are available to estimate ionospheric content and ionospheric Faraday rotation using Global Positioning System observations and models of the Earth's magnetic field. Examples of these include \textsc{ionFR} \citep{SotomayorBeltran:2013v552p58S}, \textsc{RMextract}\footnote{\url{http://github.com/maaijke/RMextract}} and \textsc{albus} (Willis, submitted). While these tools struggle to predict fine spatial and temporal detail in the ionosphere, they provide sufficient resolution to correct for the mean electron content of the ionosphere.

\begin{figure}[ht]
\includegraphics[width=\linewidth]{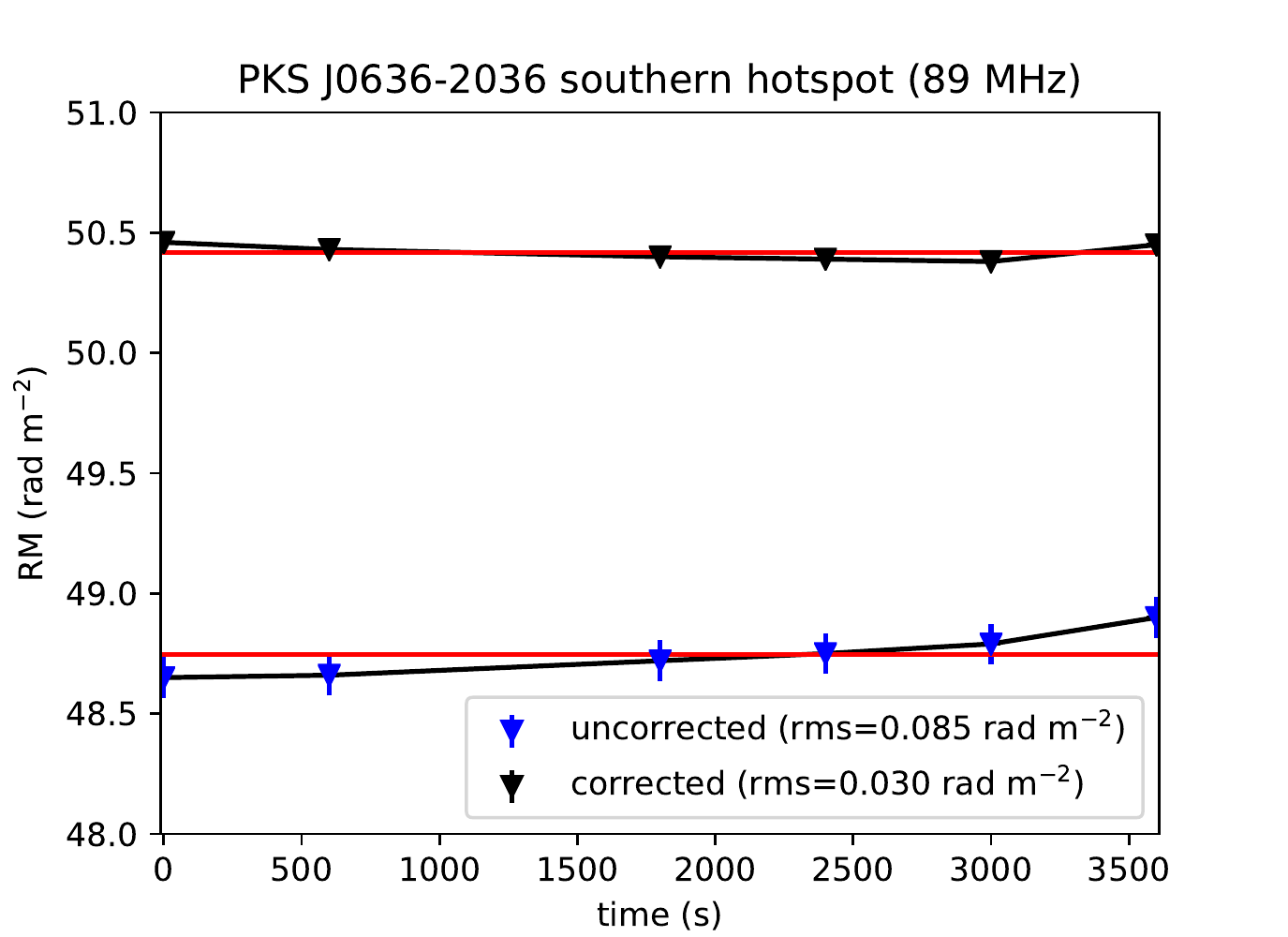}
\caption{Variation in observed rotation measure in the southern hotspot of PKS J0636$-$2036 as a function of time in the lowest MWA band ($89$ MHz). Blue points show the measured RM before ionospheric calibration and the Black points show the measured RM after ionospheric calibration. The red line shows the mean value of RM in each instance.}
\label{fig:lfiono}
\end{figure}

\begin{figure}[ht]
\includegraphics[width=\linewidth]{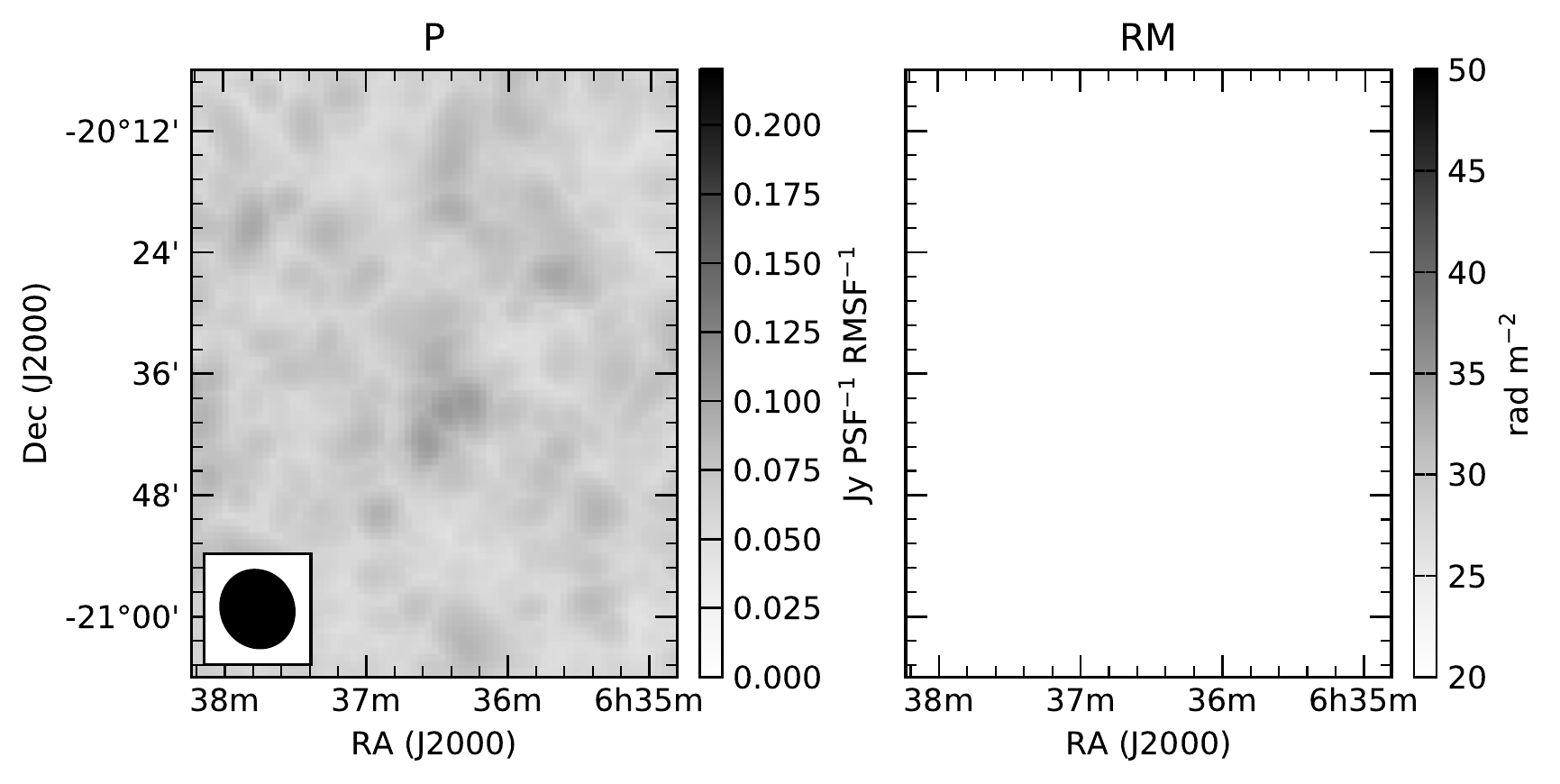}
\includegraphics[width=\linewidth]{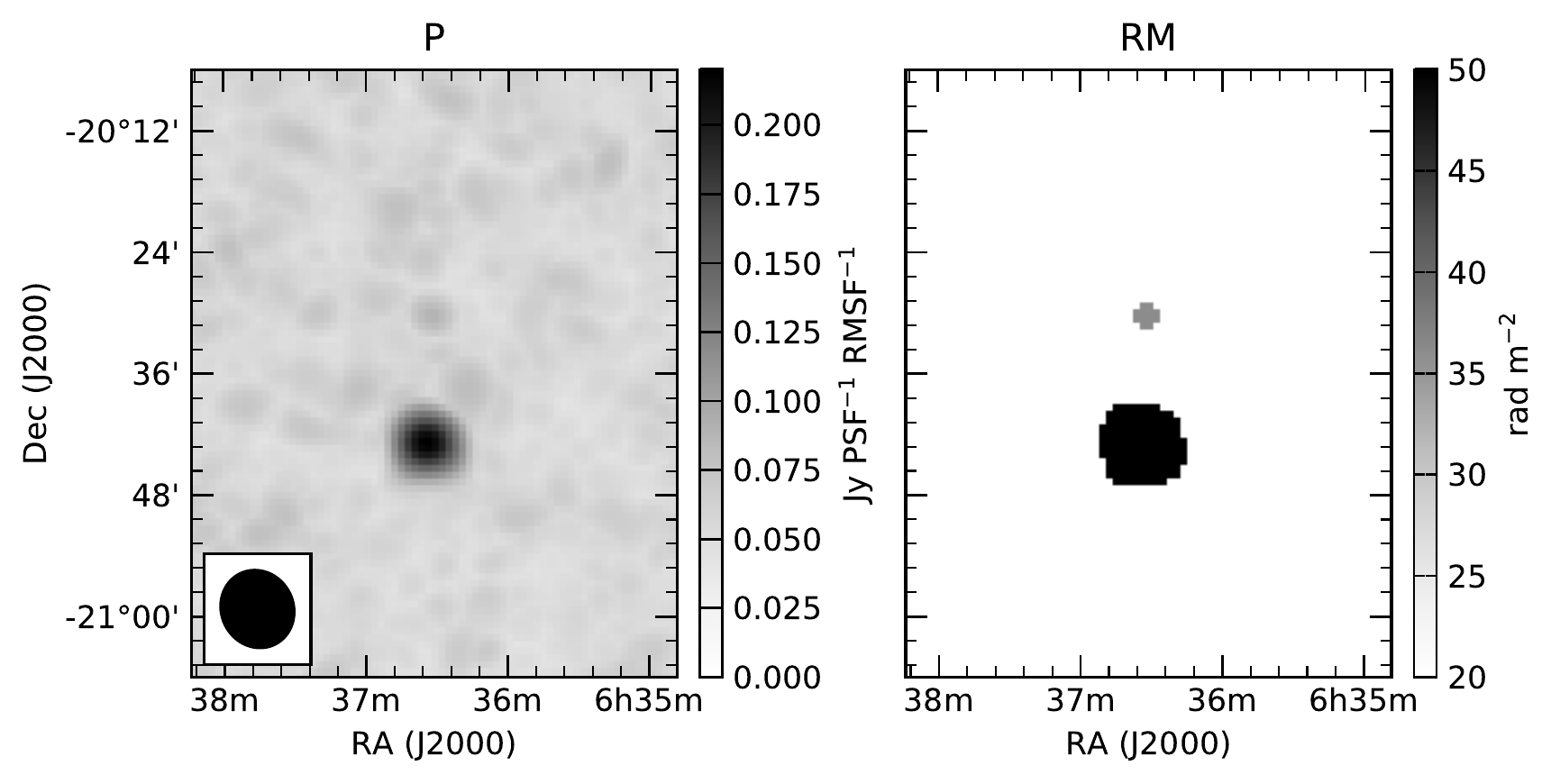}
\caption{Polarised intensity (left) and rotation measure (right) maps for PKS J0636$-$2036 in the lowest MWA band ($89$ MHz) uncorrected (top) and corrected (bottom) for ionosphere Faraday rotation. The rotation measure maps are masked to show only peaks with 6$\sigma$ significance (for a measured noise of $1\sigma=13$\,mJy\,PSF$^{-1}$ in polarised intensity).}
\label{fig:ionocor}
\end{figure}

Figure \ref{fig:lfiono} demonstrates the improvement gained by applying a correction for ionospheric Faraday rotation. The plot shows the uncorrected and corrected RM measured for polarised emission seen in the southern hotspot of PKS J0636$-$2036 as a function of time, as observed in the lowest MWA frequency band. Prior to correction, the overall RMS in RM over the duration of the observing period is 0.085 rad m$^{-2}$. As can be seen in Figure \ref{fig:ionocor}, this is sufficient to completely depolarise the signal of this otherwise bright source. However, when the RM is corrected for ionospheric effects, the RMS in RM reduces to 0.03 rad m$^{-2}$ and the source emission is recovered. In addition, the ionospheric correction has enabled the detection of a weaker northern component of this source. After correction, the measured RMS in RM is approximately 3 times higher than the expected RMS (0.01 rad m$^{-2}$), most likely as a result of the coarse resolution of the ionospheric modelling in both time and sky position.

\subsubsection{Ionospheric self-calibration}

The example shown in Figure \ref{fig:lfiono} assumed that we had no a priori information regarding the RM of the source. If there is a sufficiently bright polarised source in the field of view and the source has a predetermined RM or we can assume that it has a constant RM then there is no need to do ionospheric modelling. Instead, Stokes Q and U can be rotated on a per-snapshot basis to ensure that the measured RM of the source remains constant. In this sense, it is possible to self-calibrate against ionospheric Faraday rotation.

One benefit of this method is that GPS data is often not available until weeks after the event. A second benefit is that ionospheric modelling does not always give accurate estimates for ionospheric Faraday rotation at the source location and observation time. Ionospheric self-calibration provides accurate RM information without the need for external information, after initially correcting for the ionosphere using a model and the RM of the source is known.

The main limitation of ionospheric self-calibration is that it requires a sufficiently bright polarised source in the observed field of view. However, as discussed in Section \ref{sec:xyphase}, diffuse polarisation provides an alternate source of polarisation against which calibration can be performed.  \citet{Lenc:2016} demonstrated that this diffuse polarised emission could not only be used to track the effects of ionospheric Faraday rotation as a function of time (on 2-minute time-scales) but also spatially over the entire MWA field of view. 

\subsubsection{Mapping the ionosphere}

The use of diffuse polarisation to perform self-calibration for ionospheric Faraday rotation presents an interesting possibility. Assuming that a Galactic RM map can be determined, which is also one of the goals of the SKA higher frequency RM Grid experiment \citep[e.g.][and references therein]{Johnston-Hollitt:2015}, this map can be used as a template against which the ionospheric contribution can be mapped. Such a map could be used to determine the total-electron content (TEC) between the observer and the diffuse Galactic ISM, given a model for the geomagnetic field. This complements existing ionospheric mapping capabilities demonstrated with the MWA that measured spatial gradients in TEC \citep{Loi:2015v42p3707,Loi:2015v50p574,Loi:2015v453p2731} as opposed to absolute TEC. It would also enable wide-field studies of the ionosphere over finer temporal and spatial scales than is currently possible with ionospheric modelling tools.

\subsubsection{Validating polarisation with the ionosphere}

A side effect of polarisation leakage is that it manifests itself as a peak at RM$\approx$0\,rad\,m$^{-2}$; an effect observed in wide-field polarimetric observations with parabolic dishes at shorter wavelengths e.g. \citet{Jagannathan:2017}. This can complicate the detection of real polarised sources with low RM. However, \citet{Lenc:2016} showed that the ionosphere can be used as a tool to detect such sources. Using multiple observations of the source over multiple epochs where the ionospheric conditions vary significantly can aid to push the measured RM of the source away from an RM of 0 rad m$^{-2}$ allowing a separation of the source RM from the peaks caused by polarisation leakage.

Ionospheric Faraday rotation can also be used to validate polarised source candidates. By cross-correlating the Faraday dispersion functions of a source over multiple epochs, if the observed shift in source RM matches the shift expected by the ionosphere then this provides greater confidence that the candidate source is real.

\subsection{Remaining calibration challenges}
\label{sec:challenges}

The regular spacing of dipoles in a MWA tile results in a primary beam with significant primary beam grating lobes. Bright sources located in these gratings generate PSF sidelobe structures that sweep through the main beam as a function of frequency and time. These sidelobe structures can appear highly polarised if they originate from low elevations where the beam has a different polarised response, and deconvolution will be limited because the polarised response for such sources is not well modelled. This can result in false detections in Faraday space and limit the detection of weakly polarised sources.

For bright point sources, ``peeling'' can be used to minimize the PSF sidelobe structure from a problematic source \citep{Mitchell:2008v2p707}. Peeling refers to the $(u,v)$-plane subtraction of the source model multiplied by its best fit Jones matrices so as to remove the source contribution from all four instrumental polarisations. For unresolved sources this is difficult to achieve because a well defined model is required. Furthermore, at low elevation, the projected baselines of the MWA are severely foreshortened making the array sensitive to extended sources \citep{Thyagarajan:2015}. In particular, the presence of the Galactic plane in one of the primary beam grating lobes can be detrimental for imaging. The methods for dealing with such cases are limited, typically requiring avoiding beamformer pointings where such instances occur. Alternatively, an adaptation of SAGECal for use with MWA observations may be used to remove PSF sidelobe structures using direction dependent calibration \citep{Yatawatta:2008,Kazemi:2011}. Further investigation is required to solve this problem.

Another remaining challenge is calibrating for the absolute polarisation angle. Currently, the polarisation angle is left unconstrained during the calibration procedure. This is reasonable for polarimetry over a single observing session but may lead to depolarisation when integrating over multiple observing sessions if the polarisation angle is inconsistent in each session. In practice, absolute polarisation angle calibration is complicated by beamformer pointing changes and ionospheric Faraday rotation and the solution to this problem requires further development. Especially when there are little or no reference measurements at low frequencies, and an artificial ‘cal’ signal cannot be injected into the receivers, as in single dish observations, e.g., \citet{Johnston:2005}.

\section{Future directions for polarimetric science}
\label{sec:science}

Our experience with MWA has laid the foundations for future work in the study of polarimetric science. In this section we discuss some of the outstanding questions that polarimetric observations can give us insight into. 

\subsection{Low frequency polarisation source counts}
\label{sec:mwapoint}

Linear polarisation source counts are well studied at 1.4 GHz \citep[e.g.][]{Taylor:2007, Hales:2014,Rudnick:2014} and at 20 GHz \citep{Massardi:2013}. However, at low frequencies there have been only a small number of detections to date. 
Two surveys with the MWA detected only two extragalactic sources brighter than 300 mJy\,PSF$^{-1}$ (in Section \ref{sec:compact}; a shallow survey over $\sim$6000 square degrees) and four 30$-$300 mJy\,PSF$^{-1}$ in a deeper survey over $\sim$400 square degrees \citep{Lenc:2016}. A pointed observation with LOFAR, with a resolution of $20\arcsec$, detected one source every $\sim$2 square degrees for sources between 0.5$-$6.0 mJy PSF$^{-1}$ in a $\sim$17 square degree region \citep{Mulcahy:2014}. \citet{Jelic:2015} also detected 15 polarised sources within a $\sim$25 square degree region towards 3C196; however, the details (location, flux density, polarisation fraction, and RM) of the sources were not presented.

MWA observations can provide an improved understanding of the polarised source population above $\sim$10 mJy\,PSF$^{-1}$ over the entire sky because of the wide-field capabilities of the instrument. If the findings of \citet{Lenc:2016} hold true over the entire sky then of order $\sim 250$ polarised extragalactic sources should be detected. At this level, sources counts will become more statistically relevant.

As described in Section \ref{sec:compact}, a significant factor that limits the effectiveness of point-source polarimetry with the MWA is its limited angular resolution. The $2\arcmin-4\arcmin$ synthesised beam of the MWA results in significant beam depolarisation in all but the simplest of sources. This may improve in future as the MWA project considers extended baseline lengths. Such an extension will have a two-fold improvement as it will not only reduce beam depolarisation effects but also improve sensitivity in imaging modes that use uniform weighting as the longest baselines will no longer be significantly down-weighted.

\subsection{Low frequency depolarisation of radio galaxies}

Low frequency linear polarisation observations of extragalactic radio sources can enable detailed studies of magnetised plasma along the entire line of sight from the emitting source to the telescope. The very large wavelength-squared coverage of low frequency telescopes means they can outperform traditional centimetre radio facilities by more than two orders of magnitude in terms of the RM resolution ($<1$~rad~m$^{-2}$) of the observations (Table \ref{table:polpar}). However, finding brightly polarised extragalactic sources at low frequencies has been challenging due to the often poor angular resolution and the strong effect of Faraday depolarisation \citep[e.g.][]{Farnsworth:2011}. Therefore, obtaining a better understanding of exactly how radio galaxies depolarise at low frequencies is important for designing experiments to detect a sufficiently large sample that would allow statistical studies of the emitting sources themselves as well as the foreground magnetoionic material \citep[e.g.][]{Gaensler:2015,Vacca:2015}. 

There are multiple models describing the depolarisation behaviour of extragalactic sources as a function of wavelength. The predictions made by these models tend to diverge at long wavelengths e.g. an exponential fall-off \citep[or the ``Burn-law''][]{Burn:1966}, a power-law fall-off \citep[the ``Tribble-law''][]{Tribble:1991, Tribble:1992}, and a depolarisation curve with a non-monotonic behaviour at long wavelengths \citep{Sokoloff:1998}. Thus, low-frequency polarisation observations provide a new parameter space for efforts to determine the true physical nature of the magnetionic media causing the observed Faraday depolarisation.  Targeted searches with the MWA for polarised emission from large angular-size and linear-size radio galaxies in poor environments, such as in PKS~J0636$-$2036 (O'Sullivan et al.~2017, in prep), will likely yield sources for which depolarisation studies can be conducted. 

\subsection{Pulsars}
\label{sec:pulsars}
Polarisation observations of pulsars allow us to probe their magnetospheres and the intervening plasma, and may be an alternate way to find new pulsars \citep{Gaensler:1998,Navarro:1995}. The MWA has already been used to study the integrated properties of pulsars \citep[e.g.,][]{Bell:2016,Murphy:2017} and their high-time resolution properties \citep[e.g.][]{Bhat:2016v818p86,McSweeney:2017}.

Phase-resolved observations of pulsars show signals that can be up to 100\% polarised, with prominent linear polarisation showing a changing position-angle across the pulse phase and significant circular polarisation as well (e.g., \citealt{Gould:1998, Lorimer:2012}).  The linear polarisation swings are often interpreted in the context of the ``rotating vector model'' \citep{Radhakrishnan:1969}, which allows determination of the beam size, inclination, and emission altitude through multi-frequency studies. Even phase-averaged observations of pulsars can show significant polarisation, and this is borne out by the propensity of pulsars to dominate the types of compact polarised sources detected with the MWA (see Sections \ref{sec:compact} and \ref{sec:circpol}).

Measuring dispersion measure (DM) along with RM for pulsars provides an efficient method to determine the average strength and direction (parallel to the line of sight) of magnetic fields in foreground regions, including the three-dimensional structure of the Galactic magnetic field \citep[e.g.][]{Noutsos:2008}, complementing direct probes of the ISM (Section \ref{sec:ism}). Low-frequency radio telescopes like the MWA with large fractional bandwidths can determine DMs and RMs with high precision, \citep[e.g.][]{Noutsos:2015}, potentially expanding this technique to probe, for example, globular clusters, the heliosphere (see Section \ref{sub:solar}), the ionosphere, and temporal and spatial variations within these plasmas via monitoring observations \citep[e.g.,][]{Howard:2016}.

The polarised and time-variable nature of pulsar emission is one way of identifying pulsars in the image plane. This could be used to search for unusual pulsars \citep[e.g.][]{Lenc:2016,Bell:2016} that are not easily detected in traditional periodicity searches. For example, pulsars in fast binary orbits \citep{Gaensler:1998}, that are highly scattered, or that display variability due to scintillation, magnetospheric emission phenomena \citep[nulling, mode-changing, intermittency, e.g.,][]{Sobey:2015}, or dramatic magnetic field reconfigurations in magnetars \citep[e.g.,][]{Levin:2012}. 

The noise in Stokes I MWA continuum images is dominated by confusion, with values of about 100\,mJy\,PSF$^{-1}$ in a 2\,min image \citep{Franzen:2016}.  In contrast, the noise in Stokes V is purely thermal noise-dominated, close to 20\,mJy\,PSF$^{-1}$ \citep{Tingay:2013}: a reduction of a factor of 5. With typical phase-averaged polarisation fractions of $\approx 10$\% \citep{Gould:1998,Han:98} the signal-to-noise for some pulsars could be larger in circular polarisation than in total intensity. 

\subsection{The flare rate of low-mass stars}

Flaring activity is a common characteristic of magnetically active stars. In the 1960s - 1970s several magnetically active M dwarf stars were observed at frequencies between $90 - 300$~MHz using single dish telescopes. These observations revealed bright radio flares with rates between $0.03 - 0.8$ flares per hour, intensities ranging from $0.8$ to $20$\,Jy, and high fractional circular polarisation \citep[e.g][]{Spangler:1974, Spangler:1976, Nelson:1979}. Yet recent searches for transients at low frequencies have resulted in non-detections \citep[e.g][]{Rowlinson:2016, Crosley:2016}. 
These non-detections call into question the M dwarf star flare rates and flux densities reported by earlier authors.  

To confirm the previous M dwarf star flare rates and flux densities at $100 - 200$~MHz, \citet{Lynch:2017b} targeted the magnetically active M dwarf star UV Ceti. Taking advantage of the expected high fractional circular polarisation for any detected flares, they focused their transient search in Stokes V images rather than Stokes I. \citet{Lynch:2017b} detected 4 flares from UV Ceti, with flux densities a factor of 100 fainter than those reported  previously in the literature. These faint flares were only detected in the polarised images, which had an order of magnitude better sensitivity than the total intensity images. This highlights the utility of polarised imaging to detect low-level transient emission for confusion limited radio telescope arrays.  

\subsection{Direct detection of exoplanets}

Planets with strong magnetic fields can produce radio emission when energetic electrons propagate along these fields towards the surface of the planet. The magnetised planets in our own solar system are known to emit intense radio emission from their auroral regions through the electron-cyclotron maser instability (CMI) \citep[e.g.][]{Zarka:2001}. Analogously, magnetised exoplanets are expected to be sources of low-frequency radio emission.  Like the emission from flaring M dwarf stars, the radio emission observed from magnetised planets has high fractional circular polarisation. Two studies using Stokes V imaging with the MWA have targeted known exoplanets \citep{Murphy:2015v446p2560} and theoretical best candidates for radio detections \citep{Lynch:2017a}; both resulted in non-detections.  

\subsection{The structure of the local ISM}
\label{sec:ism}

Recent long wavelength polarimetric observations of the diffuse ISM have provided the first glimpses of the structure of the local ISM \citep{Jelic:2015,Lenc:2016,VanEck:2017}. These have only covered small regions of the sky, but, as shown in Figure \ref{fig:diffuse}, there is potential to use the MWA to map the entire sky for diffuse polarisation. Increasing the area imaged for diffuse polarisation should reveal large local structures, associated with turbulence or perhaps even supernova remnants, that are not easily seen using other wavelengths.

Maps of polarised diffuse emission are of particular interest for EoR projects using facilities such as the Hydrogen Epoch of Reionization Array \citep[HERA,][]{DeBoer:2017}, Experiment to Detect the Global EoR Step \citep[EDGES,][]{Bowman:2010v468p796}, Dark Ages Radio Explorer \citep[DARE,][]{Burns:2012v49p433}, and the Square Kilometre Array \citep[SKA,][]{Pritchard:2015}. Firstly, they provide input to enable removal of any potential contamination of these polarised foregrounds from observations of EoR fields. Secondly, they enable efforts to find regions of the sky where such foregrounds are minimal. Observing polarised diffuse emission at $<100$\,MHz is also of interest to Cosmic Dawn experiments in order to map potential contaminants. 

The existence of diffuse polarised emission poses challenges for calibrating and combining EoR datasets. While a static polarised background can be measured and corrected in each dataset, Faraday rotation due to ionospheric activity will change the spectral characteristics and the polarised emission, yielding a time-dependent signal in Stokes I when polarisation purity is not high. The EoR experiment relies on the spectral smoothness of foregrounds across frequency to discriminate cosmological signal from contamination. Non-zero rotation measures may imprint spectral structure, which can leak into the parts of parameter space desired to be free of foregrounds \citep{Geil:2011}. In addition, the existence of diffuse polarised emission, with spatial structure on scales of the EoR signal, can mimic real cosmological signal, further complicating signal discrimination. The mitigation of polarisation leakage into Stokes I and estimating the effect of this leakage on EoR science are areas of ongoing research; see \citet{Jelic:2010}, \citet{Asad:2016}, \citet{Moore:2017}, and references therein.

\subsection{Magnetic fields in the intergalactic medium}

Upgrades to the MWA may provide the capabilities needed to directly detect magnetic fields in the intergalactic medium (IGM), often called the ``cosmic web''. The IGM is the filamentary web of plasma existing between galactic structures, and occupies the vast majority of the volume of the Universe \citep{Cisewski:2014}. Magnetic fields in this plasma have likely moderated the large-scale structure of the Universe. However, the diffuse nature of the plasma and the theoretically predicted few-nG magnetic fields have conspired such that IGM magnetism currently remains completely undetected. Using the MWA and newly developed techniques, it may be possible to provide a statistical detection of the magnetised IGM that is not reliant on a priori knowledge of the cosmic web distribution.

Total intensity radio images from the MWA at 180~MHz have already been combined with tracers of the large-scale structure in order to place limits on the magnetic field strength \citep{Vernstrom:2017}. Furthermore, the MWA will also provide polarised low-frequency counterparts to several sources detected at higher radio frequencies which correspond to sources related to over- and under-densities (filaments and voids) in the IGM. 

A large lever-arm in wavelength-space is known to be essential to properly detect the magnetised component of the IGM \citep{Akahori:2010,Akahori:2011, Akahori:2014,Ideguchi:2014}, and the combination of MWA, GMRT, and Australia Square Kilometre Array Pathfinder (ASKAP) data should be able to provide a direct detection. The GMRT declination range for polarisation studies \citep{Farnes:2014} reaches down to $-53$~degrees, thus providing an overlapping range for MWA combined studies. The low-frequency nature of the MWA considerably improves the RM resolution in Faraday space, and makes the MWA critical for this experiment. This is estimated using the model from \citet{Akahori:2010} which assumes an IGM magnetic field of 10~nG in filaments, finding that this leads to a detectable RM excess of 1~rad~m$^{-2}$. This is consistent with other studies \citep{Vazza:2014,Vazza:2016}, which find that the typical field in filaments ranges from a few nG and up to $\sim$100~nG. Using the ultra-fine RM resolution afforded by these ultra-broadband observations, the magnetised IGM should be detectable with the order of 100 sources. Advantageously, we do not need to know if the source is seen through a filament or a void, however we can subsequently only measure an average magnetic field in extragalactic space. The subsequent ultra-broadband polarisation spectral energy distributions will be able to reveal the IGM's magnetic field, and the combination of data from several radio facilities will yield a dataset with a bandwidth equivalent to that from the SKA. This will provide the first statistical measurement of the average magnetic field in intergalactic space, although it will not allow for any comparative studies between filaments and voids. The success of such an ambitious statistical experiment relies on the availability of high-quality models of the Galactic RM contribution which will be readily available in the near future thanks to higher frequency surveys such as the Polarization Sky Survey of the Universe's Magnetism (POSSUM) with ASKAP \citep{2010AAS...21547013G}.

\subsection{Understanding the solar environment}
\label{sub:solar}

MWA polarimetry can also provide information about the solar atmosphere and space weather events. For direct solar imaging, circular polarisation has traditionally been most relevant because large Faraday rotations 
in the low corona ($\lesssim 2~\rm{R}_{\odot}$) make linear polarisations difficult to detect due to bandwidth depolarisation (e.g. \citealt{Suzuki:1985,Gibson:2016}). Circularly polarised bremsstrahlung emission is produced continuously in active regions, where it encodes information about the temperature gradient and magnetic field \citep{White:1999,Grebinskij:2000}.

Solar radio bursts also exhibit circular polarisation, but standard coherent emission theories predict much higher fractions than are observed (e.g. \citealt{Cairns:2011,Reid:2014}). This discrepancy has not been satisfactorily resolved, and the MWA's ability to localise burst emission can contribute significantly. For instance, the depolarising effects of scattering \citep{Melrose:1989} and linear mode conversion \citep{Kim:2007} may produce observable spatiotemporal variations in the polarisation structure of particular type III bursts and/or a statistical bias toward lower polarisations closer to the limb.

Linear polarisations can be detected in solar images of sufficiently high spectral resolution \citep{Segre:2001}, which has been accomplished for microwave observations with $\sim$20 kHz spectral resolution \citep{Alissandrakis:1994}. The  10 kHz spectral resolution of the MWA may facilitate analogous low-frequency studies of the magnetic field strength above active regions. Linearly polarised background sources viewed through the upper corona have also been 
detected \citep{Spangler:2007,Ingleby:2007,Ord:2007}; similar observations would be challenging for a widefield imager like the MWA but may be possible given recent advances in the achievable dynamic range. 

Further from the Sun, the magnetic field weakens and Faraday rotation signatures in the interplanetary medium become detectable at low frequencies \citep{Oberoi:2012}. Of particular interest is the magnetic field orientation in front of and inside Coronal Mass Ejections (CMEs), since Earth-bound CMEs can cause dramatic space weather events that depend strongly on the time-varying magnetic field orientation. Measurements of changes in Faraday Rotation of background sources during the passage of a CME have been made \citep[e.g.][]{Howard:2016,Kooi:2017}, but the interpretation at low frequencies is hampered by uncertainties in the ionospheric contribution to the observed Faraday rotation and by the sparsity of polarised sources. The ubiquity of the diffuse Galactic polarised emission means that it may be possible to map the Faraday rotation signature of a CME in detail. Such an imaging approach would also make it easier to separate the effects of the ionosphere, since changes in the ionospheric FR across the field of view are likely to be small, and since structures in the ionosphere would be unlikely to match the CME in morphology or velocity.

Additionally, \citet{Bastian:2001} have shown that synchrotron radiation may be emitted by CMEs soon after launch. This may offer an alternative route to determining the CME magnetic field if linearly polarised emission can be detected.

\section{Conclusion}

Our work with the MWA forms a foundation for future work in polarimetry at low frequencies leading to SKA-LOW. One valuable lesson is that understanding the instrument beam response and the ionosphere are key. In many ways, the overall design of the MWA has helped greatly in our understanding of both of these. For example, with over 8000 baselines, the MWA has excellent instantaneous $(u,v)$-coverage. This provides a snapshot imaging capability, which has not only been a great enabler for transient science but has also provided a means to sample the instrumental beam and the ionosphere on fine time-scales.

The compact and dense baseline configuration of the MWA also has enabled imaging of diffuse linear polarisation from the local ISM. This resulted in alternate approaches to polarimetric calibration and correcting for ionospheric Faraday rotation. While the compact configuration of the MWA has limited the resolution of the instrument, thereby increasing the detrimental effect of beam depolarisation, it has simplified ionospheric calibration by ensuring the MWA is in the regime where each tile sees through the same ionospheric patch. Our work shows that beam depolarisation has not prevented us from detecting and studying polarised sources.

The MWA has gained much by starting small and progressively increasing complexity, incorporating incremental improvements along the way. The instrument is currently undergoing upgrades to utilise redundant baselines and to increase baseline lengths. The lessons gained in performing MWA polarimetry are also valuable lessons for SKA-LOW.  SKA-LOW will also need to correct for the ionosphere and will need to understand its beam model. To do so effectively may require careful consideration of how stations are arranged and the availability of short baselines.

\begin{acknowledgements}
The Centre for All-sky Astrophysics (an Australian Research Council Centre of Excellence funded by grant CE110001020) supported this work. This scientific work makes use of the Murchison Radio- astronomy Observatory, operated by CSIRO. We acknowledge the Wajarri Yamatji people as the traditional owners of the Observatory site. Support for the operation of the MWA is provided by the Australian Government (NCRIS), under a contract to Curtin University administered by Astronomy Australia Limited. We acknowledge the Pawsey Supercomputing Centre which is supported by the Western Australian and Australian Governments. We acknowledge the International Centre for Radio Astronomy Research (ICRAR), a Joint Venture of Curtin University and The University of Western Australia, funded by the Western Australian State government. The Dunlap Institute is funded through an endowment established by the David Dunlap family and the University of Toronto. B.M.G. acknowledges the support of the Natural Sciences and Engineering Research Council of Canada (NSERC) through grant RGPIN-2015-05948, and of the Canada Research Chairs program. The authors thank the anonymous referee for providing useful comments on the original version of this paper.
\end{acknowledgements}

\bibliographystyle{pasa-mnras}
\bibliography{biblio}

\end{document}